\pdfoutput=1 

\documentclass[useAMS,usenatbib,aas_macros]{mn2e}

\usepackage{amssymb,amsmath}
\usepackage{graphicx}
\usepackage{flafter}
\usepackage{placeins}
\usepackage{subfigure}
\usepackage{booktabs}
\usepackage{array}
\usepackage{natbib}
\usepackage{aas_macros}
\usepackage{multirow}
\usepackage[T1]{fontenc} 
\usepackage{aecompl}
\title[Effects of B/P Bulges on Galaxy Models]{The Effects of Boxy/Peanut Bulges on Galaxy Models}
\author[Fragkoudi et al.]{F. Fragkoudi$^{1}$\thanks{E-mail:
francesca.fragkoudi@lam.fr}, E. Athanassoula$^{1}$, A. Bosma$^{1}$, F. Iannuzzi$^{1}$  \\
$^{1}$Aix Marseille Universit\'{e}, CNRS, LAM (Laboratoire d'Astrophysique de Marseille) UMR 7326, 13388, Marseille, France \\}
\begin{document}

\date{}

\pagerange{\pageref{firstpage}--\pageref{lastpage}} \pubyear{2014}

\maketitle

\label{firstpage}

\begin{abstract}
	
We examine the effects that the modelling of a Boxy/Peanut (B/P) bulge will have on the estimates of the stellar gravitational potential, forces, orbital structure and bar strength of barred galaxies. 
We present a method for obtaining the potential of disc galaxies from surface density images, assuming a vertical density distribution (height function), which is let to vary with position, thus enabling it to represent the geometry of a B/P. We construct a B/P height function after the results from a high-resolution, $N$-body+SPH simulation of an isolated galaxy and compare the resulting dynamical model to those obtained with the commonly used, position-independent ``flat'' height functions. We show that methods that do not allow for a B/P can induce errors in the forces in the bar region of up to 40\% and demonstrate that this has a significant impact on the orbital structure of the model, which in turn determines its kinematics and morphology. Furthermore, we show that the bar strength is reduced in the presence of a B/P. 
We conclude that neglecting the vertical extent of a B/P can introduce considerable errors in the dynamical modelling.
We also examine the errors introduced in the model due to uncertainties in the parameters of the B/P and show that even for generous but realistic values of the uncertainties, the error will be noticeably less than that of not modelling a B/P bulge at all.
\end{abstract}



\section{Introduction}

Many edge-on disc galaxies can be seen to contain boxy-, peanut- or X-shaped isophotes, which are usually grouped together into the category of Boxy/Peanut (hereafter B/P) bulges. As a result of a number of theoretical studies (see \citealt{Athanassoula2008, Athanassoula2015} for general reviews of this subject), including orbital structure and stability analysis \citep{Binney1981,Pfenniger1984a,Pfenniger1985,Patsisetal2002,Skokosetal2002, Skokosetal2002b, MartinezValpuestaetal2006, HarsoulaKalapotharakos2009, ContopoulosHarsoula2013}, and numerical simulations \citep{CombesSanders1981, Combesetal1990, Rahaetal1991, Mihosetal1995, AthanassoulaMisiriotis2002, Athanassoula2003, Athanassoula2005, Bureau&Athanassoula2005, MartinezValpuestaetal2006}, these structures are now known to be due to vertical instabilities in the bar, which cause it to `puff up', giving rise to boxy or peanut-like shapes. These studies also show that once a bar forms, a B/P bulge will form soon after.

Observational studies have further confirmed the link between B/P bulges and bars (see \citealt{KormendyKennicutt2004} and \citealt{Kormendy2015} for reviews on the subject), by showing that the fraction of edge-on disc galaxies with B/P bulges is comparable to the fraction of disc galaxies containing bars \citep*{Luttickeetal2000}. Kinematic studies of edge-on barred galaxies and B/P bulges also confirm the connection between the two structures \citep[and references therein]{Athanassoula&Bureau1999, Bureau&Athanassoula1999, Chung&Bureau2004, Bureau&Athanassoula2005}.
Therefore, barred galaxies at present and past epochs will contain B/P bulges, and in fact, one is also believed to be present in our own Galaxy (\citealt{Weilandetal1994, Howardetal2009, Shenetal2010,McWilliamandZoccali2010}, Ness et al. 2012, 2013a,b; \nocite{Nessetal2012,Nessetal2013b,Nessetal2013a } \citealt{Vasquezetal2013,WeggGerhard2013,Gardneretal2014, Netafetal2013, Netafetal2014, Netafetal2015}).

Bars are found in about two thirds of disc galaxies in the local universe, with variable strengths \citep{Eskridgeetal2000,Menendezetal2007,Barazzaetal2008,Aguerrietal2009,Gadotti2009}, and are known to be the main drivers of the secular phase of galaxy evolution. The torque they induce into the disc causes outward angular momentum transfer, which in turn will cause a redistribution of matter in the disc. They are thus responsible for driving gas to the centre of their host galaxy \citep{Athanassoula1992b}, forming discy pseudo-bulges \citep{KormendyKennicutt2004,Athanassoula2005}, redistributing stars in the galactic disc \citep{Sellwood&Binney2002,Roskaretal2008,MinchevFamaey2010}, and possibly creating a fuel reservoir for AGN activity (\citealt{Shlosmanetal1990,CoelhoGadotti2011, Emsellemetal2014}, but see also \citealt{Leeetal2012}. For a review see \citealt{Combes2001}). However, even though the effect of bars on all these processes has been thoroughly examined, a study of the effects of B/P bulges on all these processes has not, until present, been carried out.

As a first step towards understanding the effect B/P bulge geometry may have on the aforementioned processes, in this paper we focus on their influence on dynamical models of their host galaxies. These models are obtained directly from images of the galaxies' surface brightness, by first assuming a vertical density distribution, or height function, and subsequently deriving the potential of the galaxy. They have been used extensively in the literature, with one of their most important implementations being in simulations that study the response of gas in a fixed potential. These response simulations are used to study the dark matter content of galaxies and to test the maximum disc hypothesis \citep{Kranz2001, Weiner2001,Slyzetal2003,Kranzetal2003, Perezetal2004}, the bar pattern speed \citep*{LLA1996,Kalapotharakosetal2010b} as well as the kinematical and morphological properties of gas in galaxies \citep{Linetal2011, Linetal2013}. Dynamical models have also been used in studies determining the bar strength \citep{ButaBlock2001,HeikiEija2002} and the orbital structure of galaxies \citep{Quillenetal1994, Patsis1997, Kalapotharakosetal2010a,Patsisetal2010}. Furthermore, they have been used to study gravitational torques in barred and spiral galaxies in order to establish the amount of gas inflow and by extension determine the importance of secular evolution \citep{Zaritsky&Lo1986,Haanetal2009,Foyleetal2010}.
In all of these aforementioned studies, the geometry of the B/P bulge is not taken into account when constructing the height function, and instead a position independent, `flat' height function is assumed. 
This is partly due to the lack of an analytical model for a B/P bulge, as well as to the inherent difficulty of detecting these bulges in face-on or intermediate inclination galaxies, which are the galaxies generally used in these studies. 

Various methods however have been proposed over the past few years, which allow either for the detection of B/P bulges, or at least for an educated guess at their existence. By viewing a large number of $N$-body+SPH simulations, and covering a wide range of viewing angles, \citet{Athanassoulaetal2014} have shown that B/P bulges manifest themselves in face-on projections as the so called `barlens' \citep{Laurikainenetal2011}, which renders their detection fairly easy. Strong observational arguments for this have been presented in \citet{Laurikainenetal2014}. Another method proposed by \citet{Debattistaetal2005}, uses signatures in the stellar kinematics of face-on or almost face-on galaxies and was implemented by \citet{MendezAbreuetal2008} who confirmed the existence of a B/P bulge in NGC 98. Furthermore, it is possible to detect signatures of B/P bulges by examining the morphological features of inclined galaxies \citep{AthanassoulaBeaton2006,ErwinDebattista2013}. 

We therefore believe that a study of the effect of B/P bulges on models of their host galaxies, and by extension of the necessity of including the geometry of B/P bulges in the height function of these models, is called for. 
To this aim, we first introduce in Section \ref{sec:section2} a straightforward and reliable method for calculating the potential, forces and derivatives of forces of a general density distribution $\rho$(r,$\phi$,z). We present some tests which demonstrate that the method can give highly accurate results, while also allowing the flexibility to choose an arbitrary height function, without being restricted to one which is constant with position. 

We then used our code on an image of an $N$-body+SPH simulated galaxy, which is presented in Section \ref{sec:smoothing}, thus obtaining a realistic potential for a barred galaxy. 
In order to create this model, we assign a thickness and a height function to the galaxy. These height functions are introduced in Section \ref{sec:section3} and include two `flat' height functions, a function which describes peanut bulges (from which we construct our fiducial B/P model), and another which describes boxy bulges.

The main results are presented in Section \ref{sec:B/P-noB/P}, where we examine the effect B/P bulges have on the potential and forces (\ref{sec:models_pea_nopea}), on the periodic orbits (\ref{sec:pea-nopea-orbits}) and on the bar strength (\ref{sec:simbarstrength}). We find that B/P bulges indeed have a significant effect on the results and therefore conclude that they should be included when modelling their host galaxy.

In Section \ref{sec:peanuterrors} we explore the errors which will be induced in the results by using a B/P model which is not exactly the `correct' one. This is necessary since it is not trivial to observationally obtain the exact parameters of B/P bulges, and this can introduce errors in the model. We show that for a range of uncertainties, the errors induced in the results are less than those induced by not modelling the B/P at all. We also introduce a new method for calculating the bar strength, $Q_T^{int}$, which takes into account the variation of the non-axisymmetric forcings along the whole extent of the bar.

In Section \ref{sec:Summary} we give a summary and list the main conclusions of our work.

\section{Method \& Tests}
\label{sec:section2}

\begin{figure}
\centering
\includegraphics[width=0.49\textwidth]{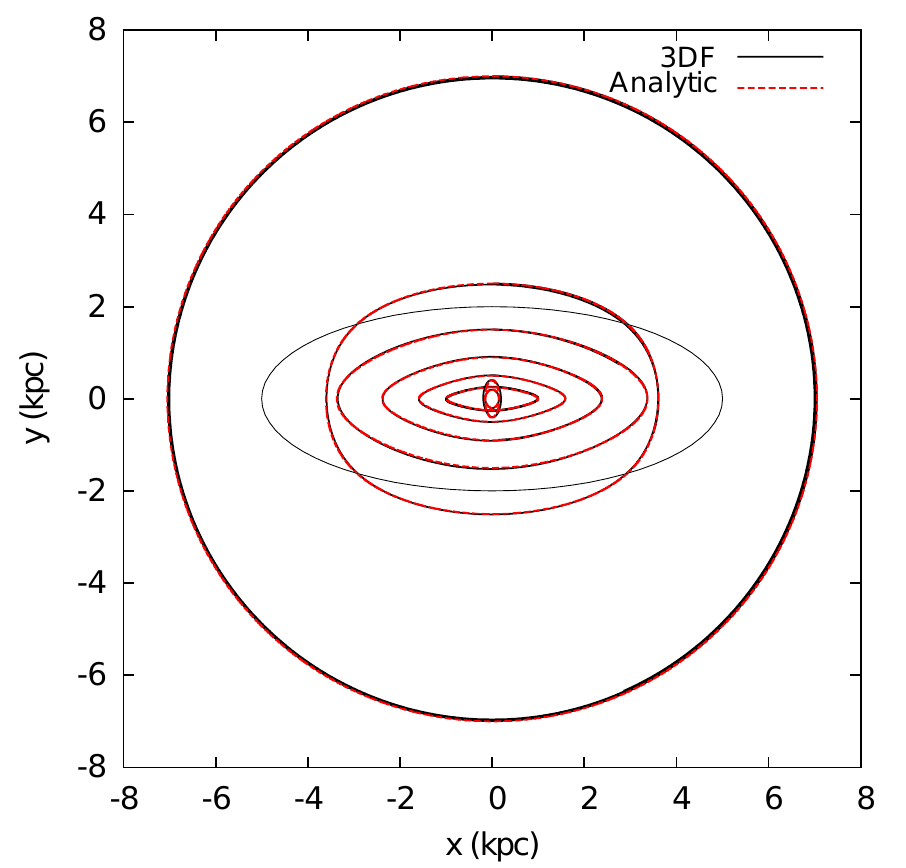} 
\caption{Comparison of orbits in the analytic and 3DF potentials of our model galaxy. The thin black line gives the outline of the bar. The orbits calculated in the 3DF potential are given in solid thick black lines, and the orbits calculated in the analytic potential are given in dashed red lines. We plot some x$_1$ orbits (along the bar), some x$_2$ orbits (perpendicular to the bar), and an almost circular orbit outside the bar region.
We see that the orbits in the two potentials almost completely overlap, so that the red and black lines are practically indistinguishable.}
\label{fig:orbits}
\end{figure}

To create a dynamical model of a galaxy we first need its density distribution. The two-dimensional surface density can be obtained from surface brightness images of a face-on disc galaxy by assuming a M/L ratio. It is important that these images are taken in a wavelength range which minimises the effect of dust extinction and traces the old stellar population (for example, the Spitzer 3.6$\,\mu$m band). In this work we use an image of a face-on simulated galaxy, and thus we do not need to account for dust extinction, nor assign a M/L ratio, as our two-dimensional image gives the surface density directly. Once we have the surface density we also need to assign a height function, and together these give us the three-dimensional density distribution, from which we can calculate the potential of the galaxy due to the stellar component. Our method for calculating the potential involves a straightforward three-dimensional integration over the density distribution and we refer to it throughout the paper as the 3DF method. We calculate the potential in Cartesian coordinates by

\begin{equation} 
\begin{split}
& \Phi (x,y,z) =\\
& -G \int_{-\infty}^{\infty} \int_{-\infty}^{\infty} \int_{-\infty}^{\infty} \frac{\rho (x', y', z')}{\sqrt{\sum_{j=1}^{3} (x_{j}'-x_{j})^{2}+\epsilon^2}} \mathrm{d}x' \mathrm{d}y' \mathrm{d}z' ,
\end{split}
\label{eq:potdef}
\end{equation}

\noindent where \emph{G} is Newton's gravitational constant, $\rho$ is the density and $\epsilon$ is the softening length which is necessary to eliminate the noise at the expense of a small bias \citep{Merritt1996,Athanassoulaetal2000}. We can differentiate the expression in Eq. \ref{eq:potdef} analytically with respect to \emph{x}, \emph{y} and \emph{z}, to obtain expressions for the the force and its derivatives. We thus rely heavily on an adequate integration algorithm, specifically one which can deal with singularities. To tackle this we use CQUAD, a doubly adaptive integration algorithm \citep{GSL}, which requires more function evaluations than other integration routines, but is more successful in dealing with difficult integrands. It computes the integral within the desired relative error limits (or precision), which the user can set.
Since we mainly work in the $z$=0 plane, we focus in what follows on the non-zero quantities in the plane: the potential $\Phi$ and the two non-zero components of the force $F_x$ and $F_y$.
As mentioned, the above, as well as what follows, concerns the potential and forces of the stellar component of the galaxy.

\subsection{Tests on the method: Relative errors}
\label{sec:relerrorstests}

In order to test the accuracy of our method, we create a model of a barred galaxy containing a disc, a bar and a classical bulge, using density distributions which have analytic solutions for the potential and forces. We then calculate the potential and forces for this model using the 3DF method, and compare the results against the analytic solutions. The general results of these tests are very positive, which demonstrates the ability of our code to deal with difficulties such as cuspy and/or non-axisymmetic density configurations. For more details on the model of the galaxy we refer the reader to Appendix \ref{sec:appendix}.

To calculate the relative errors of the potential and the derivatives of the force, we use the relation
\begin{equation}
\mathrm{Error} = 0.5\frac{|R_1 - R_2|}{|R_1|+|R_2|},
\label{eq:relativeerrors}
\end{equation}

\noindent where $R_1$ and $R_2$ are respectively the analytic and 3DF solutions.

To calculate the relative errors for the force, we use the relation
\begin{equation}
\mathrm{Error} = \frac{|F_{i1} - F_{i2}|}{\sqrt{F_{i1}^2+F_{j1}^2}},
\label{eq:relativeerrors}
\end{equation}

\noindent where $i$ and $j$ can be either the $x$ or the $y$ component of the force, and the subscripts $1$ and $2$ stand for the analytic and 3DF solutions respectively. 
We therefore normalise the error of each component of the force by the \emph{total} force at each point. This is done because our main interest in the forces is for the calculation of orbits and because on the symmetry axes of the $x$- and $y$- components of the force (in the static frame of reference), the analytic estimates of $F_x$ and $F_y$ will be equal to zero.

We stress that the precision with which the code calculates the results is an input parameter to the code. The accuracy can be as high as the user wants it to be (within the limits of machine precision), at the expense, of course, of computation time. 
We require a three dimensional integration and, due to the propagation of error at each integration, the relative precision we ask of the CQUAD algorithm for each integration has to be larger than that which we wish to achieve. Practically this means that if we ask for a relative precision of 10$^{-3}$, we will obtain a relative precision of approximately 10$^{-1}$. This is sufficient for our purposes as the error is less than 1\% for all variables. The softening is set to 10$\,\mathrm{pc}$ throughout the paper. For this precision and softening, the maximum error of the potential is 0.3\%, of $F_x$ 0.6\% and of $F_y$ 0.7\%. 

\subsection{Tests on the method: Orbits}
\label{sec:testorbits}

Even though from the results of the relative errors we see that the 3DF method gives highly accurate results, we would like to confirm that the noise in the force field does not prevent orbits from running smoothly, as they would in an analytic potential. To do this, we calculate a number of periodic orbits in the analytic potential and in the potential derived using the 3DF method for the galaxy in the frame co-rotating with the bar, in the model described in Appendix \ref{sec:appendix}. The grid used for the orbits and throughout the paper is $200 \times 200$, and the orbits are calculated using a Kick-Drift-Kick leapfrog algorithm  \citep{Hockney&Eastwood1988,Quinnetal1997,Springel2005}. 

In Fig.~\ref{fig:orbits}, we plot a few of these orbits.  In this figure and all throughout the paper the bar major axis is along the $x$-axis. In the figure we show some x$_1$ orbits, which are elongated along the bar, some x$_2$ orbits which are perpendicular to the bar, as well as some nearly circular orbits outside the bar region. We see that the orbits calculated in the 3DF potential are a very good approximation of those calculated in the analytic potential, as the two practically coincide. Thus the error which is introduced in the potential from our 3DF method and the adopted value of precision is sufficient for our purposes.


\subsection{The image}
\label{sec:smoothing}

\begin{figure}
\centering
\subfigure[IC 4290]{%
	\includegraphics[height=3.7cm]{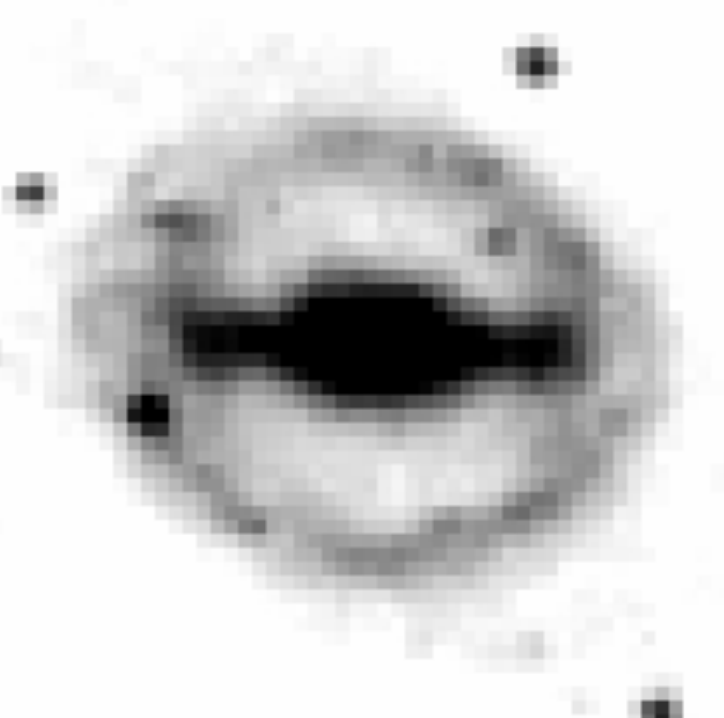}
	\label{fig:ic4290}}
\quad
\subfigure[gtr116]{%
	\includegraphics[height=3.7cm]{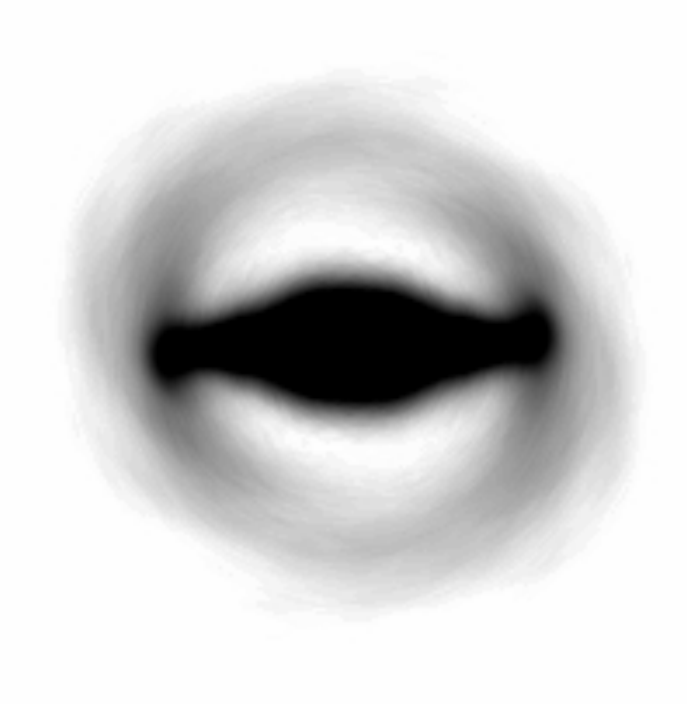}
	\label{fig:gtr116bw}}
\quad
\caption{Visual comparisson between IC 4290 and gtr116. The two galaxies have striking morphological similarities and are classified as having the same bar strength (for more details see Section \ref{sec:simbarstrength}).} 
\label{fig:IC4290}
\end{figure}

\begin{figure*}
\centering
\subfigure[Original Image]{%
	\includegraphics[width=0.24\textwidth]{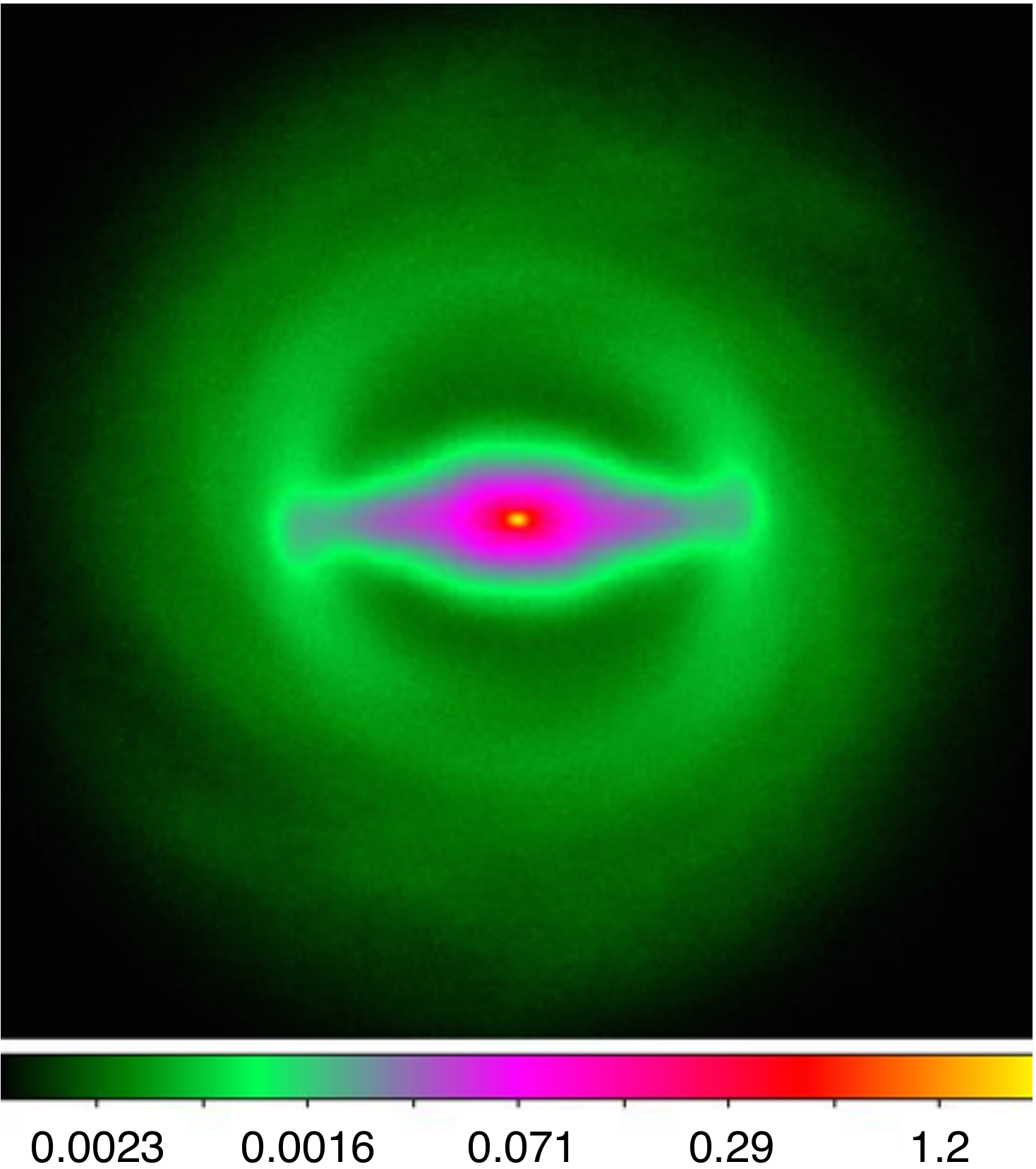}
	\label{fig:gtrorig}}
\quad
\subfigure[Model]{%
	\includegraphics[width=0.24\textwidth]{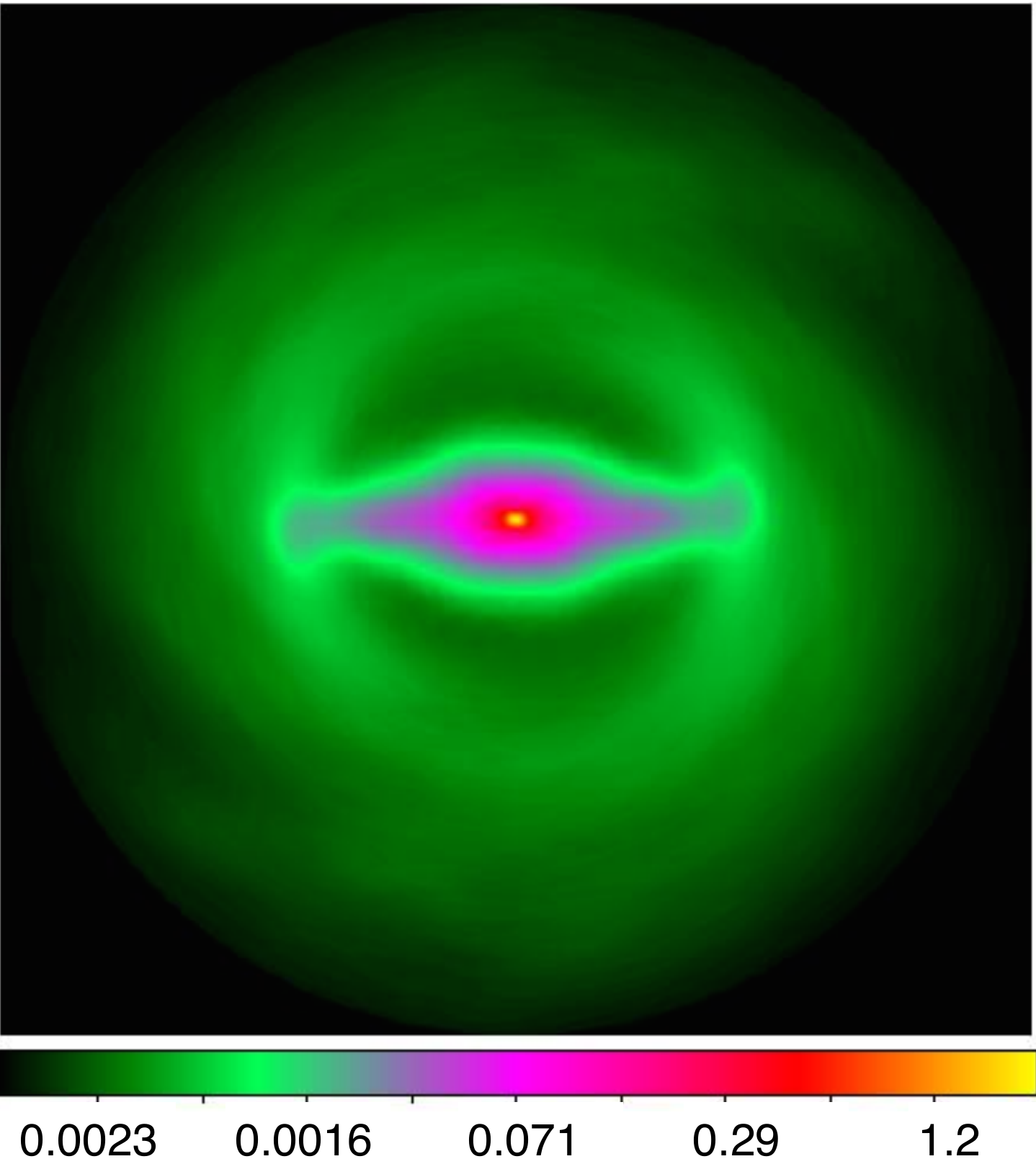}
	\label{fig:gtrrecomp}}
\quad
\subfigure[Residual]{%
	\includegraphics[width=0.24\textwidth]{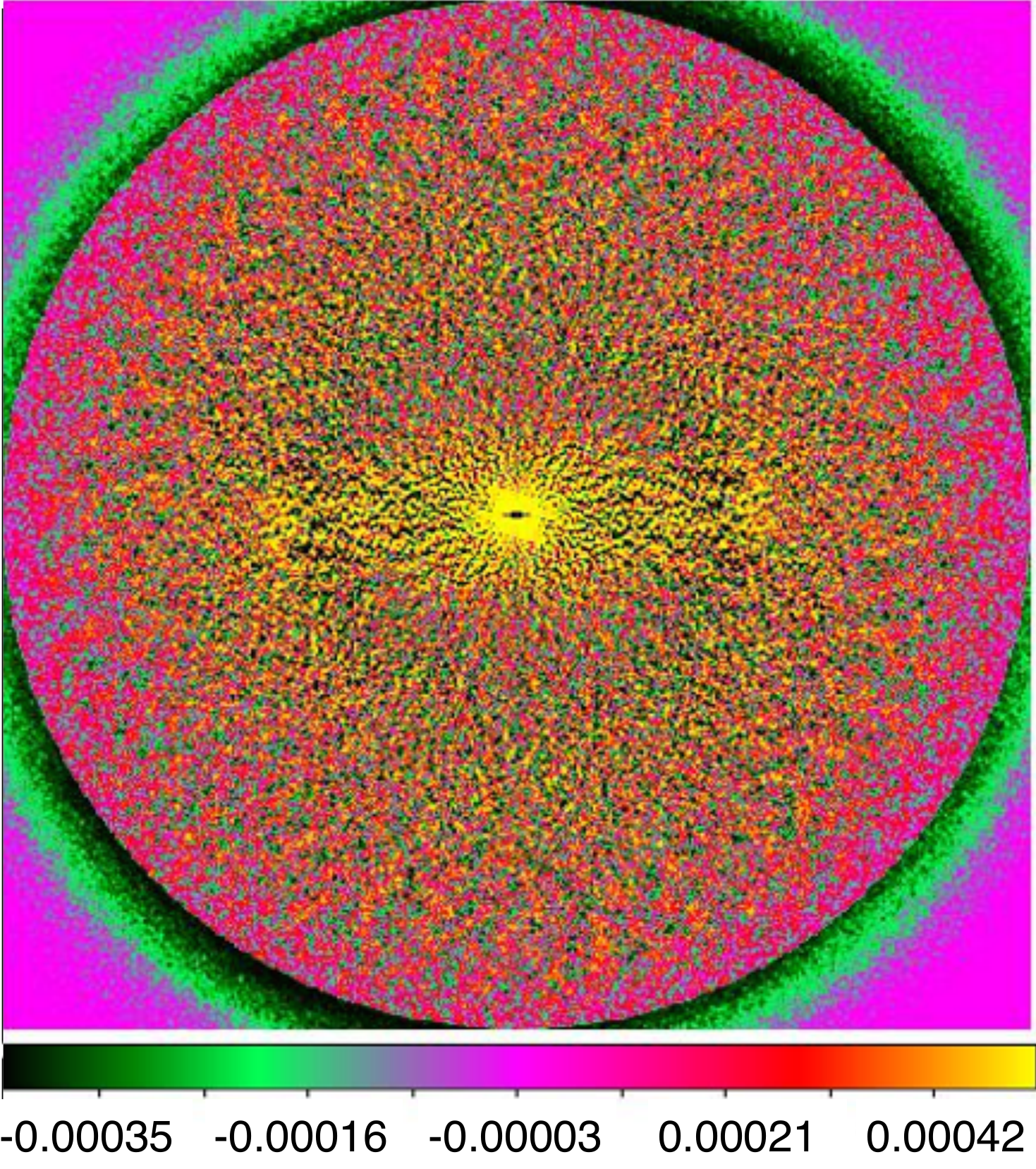}
	\label{fig:gtrresidual}}
\quad
\caption{\emph{Left}: Original image showing the surface density of the stellar component of the gtr116 simulation. \emph{Middle}: Model from the Fourier recomposition, using up to $n_F$=26 even Fourier components. \emph{Right}: Residual image after subtracting the model from the original image.} 
\label{fig:gtr}
\end{figure*}

We use the 3DF method on the density distribution derived from a face-on image of a simulated isolated galaxy and different height functions (which are described in Section \ref{sec:section3}). The initial conditions of the simulation from which the image was constructed, include a live spherical dark matter halo, an exponential stellar disc and 75\% gas (for more information on the simulation the reader is referred to run gtr116 in \citet*{AthanassoulaRodionov2013}). The snapshot we use is taken well into the secular evolution phase of the galaxy, specifically at $8\,\mathrm{Gyr}$ after the start of the simulation, by which point a strong bar and B/P bulge have formed. The image we use is constructed from the `stars' component of the snapshot and has a morphology reminiscent of that of many strongly barred galaxies, such as IC 4290 (see Fig.~\ref{fig:IC4290}). 

In order to decrease the noise of the image we require a snapshot with a large number of particles. We create a snapshot with 40 times the particles of the original snapshot, following the procedure described in \citet{Athanassoula2005}. To further reduce the noise in the image we apply some smoothing, by Fourier decomposing and recomposing it.

The Fourier components are calculated as follows:

\begin{equation}
a_n(r)=\frac{1}{\pi} \int_{-\pi}^{\pi} \! \Sigma(r,\theta) \mathrm{cos}(n\theta)\, \mathrm{d}\theta,
\end{equation}

\begin{equation}
b_n(r)=\frac{1}{\pi} \int_{-\pi}^{\pi} \! \Sigma(r,\theta) \mathrm{sin}(n\theta)\, \mathrm{d}\theta ,
\end{equation}
\noindent where $a_n$ and $b_n$ are the even and odd Fourier components, $\theta$ is the azimuthal angle, \emph{r} the radius and $\Sigma$ gives the surface density.
We then reduce the high frequency noise by recomposing the image as

\begin{equation}
\Sigma(r,\theta)=\frac{a_0}{2} + \sum\limits_{m=2}^{m=n_F} \left( a_m(r) \mathrm{cos}(m\theta) + b_m(r) \mathrm{sin}(m\theta)\right),
\end{equation}

\noindent using only a limited number of even Fourier components (in our case $n_F$=26). We show in Figs.~\ref{fig:gtrorig} and \ref{fig:gtrrecomp} the surface density of the original image and of the Fourier recomposed image, respectively, both in arbitrary units, and in Fig.~\ref{fig:gtrresidual} we show the residual image of the two. As the images of surface density are in arbitrary units, the density, as well as the potential and its derivatives will also be in arbitrary units in what follows.

\section{Height Functions Used}
\label{sec:section3}

In order to obtain the three-dimensional density of a galaxy disc from a two-dimensional image we need to assume a height function, which defines how the density drops off as a function of \emph{z} from the equatorial plane \emph{z}=0. The height function and the scaleheight ($z_0$) will of course affect the results, and we therefore need to use the height function which best approximates that of the galaxy we are trying to model. 

The height function can be either constant or can change with position. In the case where it is constant with respect to position we assume, for simplicity, that the density distribution can be written as
\begin{equation}
\rho(x,y,z)=\Sigma(x,y)F(z) ,
\end{equation}
where $\rho$ is the three-dimensional density distribution, $\Sigma$ is the two-dimensional surface density, and $F$ is the height function. In the more general case where the height function depends on position in the galaxy, as would be for example the height function describing a B/P bulge, the scaleheight changes as a function of position. In this case, the density distribution would be given by,
\begin{equation}
\rho(x,y,z)=\Sigma(x,y)F(x,y,z) ,
\end{equation}
where the normalisation of the height function is
\begin{equation}
\int_{-\infty}^\infty F(x,y,z) \mathrm{d}z=1 .
\end{equation}
It is worth noting that the mass of the model is always the same; the height function simply determines the volume density of the galaxy, by setting the thickness of the disc.
\subsection{Flat height functions}
\label{sec:flatHF}

Up to now in the literature, position-independent or `flat' height functions have been used when modelling barred disc galaxies. We therefore also use two flat height functions in this paper, to check the discrepancy which will be created in the model by a) using a flat height function and a B/P height function, and b) using two different flat height functions. We adopt two commonly used functions, the isothermal-sheet model \citep{vanderkruit1981}: 

\begin{equation}
F(z)=\frac{1}{2z_0}\mathrm{sech}^2\left(\frac{z}{z_0}\right)  ,
\label{eq:sech2}
\end{equation}

\noindent and the sech-law model \citep{vanderkruit1988}:

\begin{equation}
F(z)=\frac{1}{\pi z_0}\mathrm{sech}\left(\frac{z}{z_0}\right)  ,
\label{eq:sech}
\end{equation}
\noindent where 1/(2$z_0$) and 1/($\pi$$z_0$) are the respective normalisation factors. 
\subsection{Peanut height function}
\label{sec:PHF}

\begin{figure*}
\centering
\subfigure[gtr116: Side-on]{%
	\includegraphics[height=4.4cm]{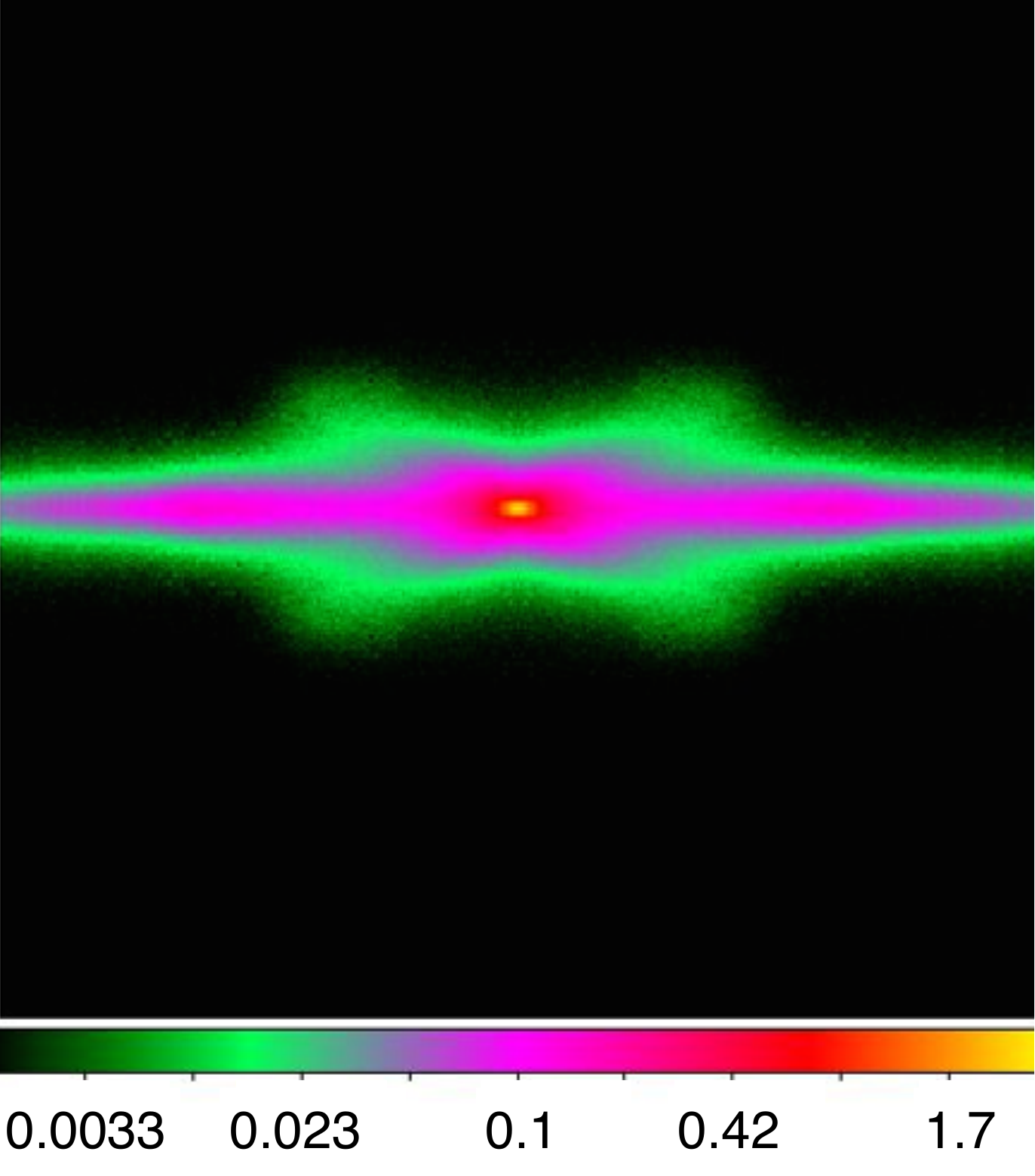}
	\label{fig:fitedge1}}
\quad
\subfigure[Edge-on: Side-on]{%
	\includegraphics[height=4.4cm]{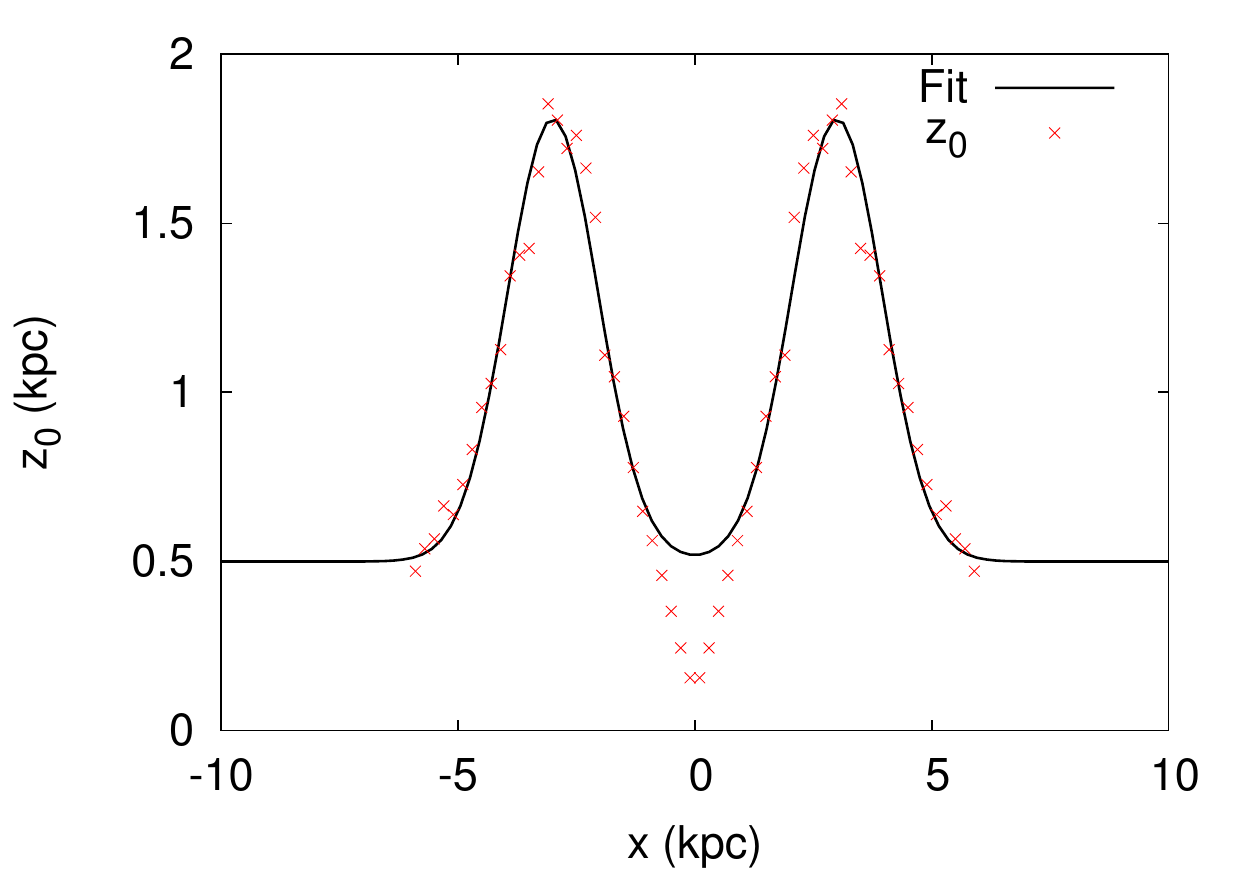}
	\label{fig:fitedge}}
\quad
\subfigure[Edge-on: End-on]{%
	\includegraphics[height=4.4cm]{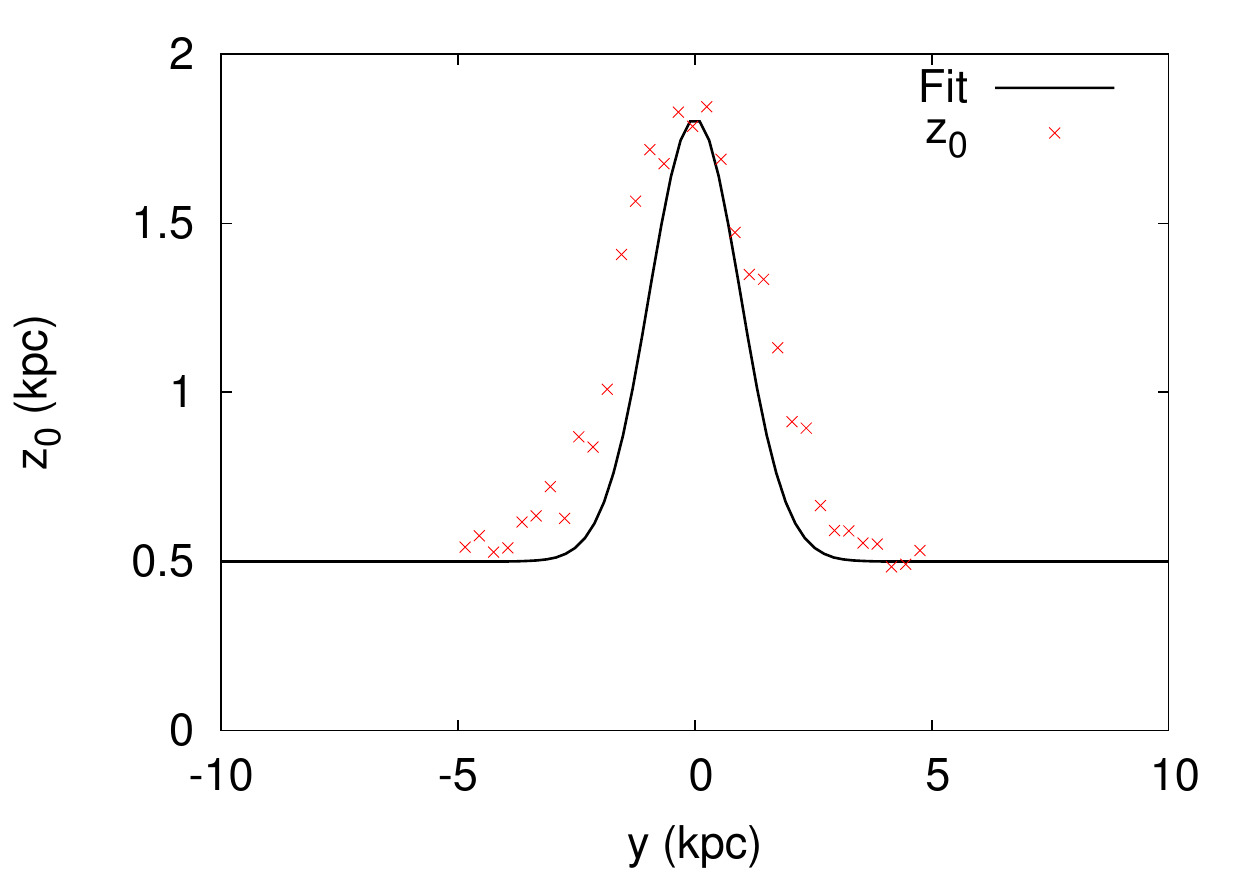}
	\label{fig:fitend}}
\quad
\caption{\emph{Left}: side-on image of the surface stellar density of the simulated galaxy gtr116. \emph{Middle}: The scaleheight of the simulation (red crosses) is plotted along the $x$-axis (for $y$=0, i.e. the side-on projection). The solid black line shows the fit of $z_0(x,0)$ to the data, which gives the scaleheight of the fiducial peanut height function. \emph{Right}: Plot of the scaleheight of the simulation (red crosses) along $x$=3 which is where the maximum of the peanut occurs (end-on projection). The solid black line shows the values of $z_0(x,y)$ along $x$=3.} 
\label{fig:fits}
\end{figure*}

To obtain a height function for the peanut, we examined the particle distribution along different cuts in \emph{x} and \emph{y} from the simulation introduced in the previous section.  We found that the sum of two two-dimensional gaussians for the scaleheight can provide a reasonable approximation to the B/P shape. As can be seen in Fig.~\ref{fig:fits} and as commented below, this choice may fail at certain points, but provides an overall fair representation of the structure.

The resulting B/P height function is a non-separable function of position and is given by:
\begin{equation}
F(x,y,z)=\frac{1}{2 z_0(x,y)}\mathrm{sech}^2\left(\frac{z}{z_0(x,y)}\right).
\label{eq:peanut1}
\end{equation}

\noindent The scaleheight \emph{z}$_0$(x,y) varies like the sum of two two-dimensional gaussians:

\begin{equation}
\begin{split}
z_0(x,y)= & A \exp\left(-\left(\frac{(x-x_0)^2}{2\sigma^2} + \frac{(y-y_0)^2}{2\sigma^2}\right)\right) + \\
& A \exp\left(-\left(\frac{(x-x_1)^2}{2\sigma^2} + \frac{(y-y_1)^2}{2\sigma^2}\right)\right) + z_0^{disc} ,
\end{split}
\label{eq:peanut2}
\end{equation}

\noindent where $A$ is the maximum scaleheight of the peanut above the disc scaleheight, $z_0^{disc}$. The variance of the gaussians is given by $\sigma^2$, ($x_0$, $y_0$) is the position of the maximum of the first gaussian and (x$_1$, $y_1$) the position of the maximum of the second gaussian.
We fit these two two-dimensional gaussians to values of the scaleheight obtained from the simulation along $y=0$ and $x=3$ (which is where the maximum of the scaleheight occurs). In the remainder of the paper, we refer to this as our fiducial peanut (or B/P) model.
 
To obtain the scaleheights, we take cuts along the \emph{x}- and \emph{y}- axes and fit the vertical particle distribution with a sech$^2$ function. We thereby determine the variation of  \emph{z}$_0$  from bin to bin along the cut. The results can be seen in Fig.~\ref{fig:fits}. In the side-on view (panel (b))  we see that the scaleheight along $y=0$ behaves approximately like the combination of two gaussians, except in the central region where the scaleheight drops below that of the outer disc. For a cut along $x=3$, where the peanut is maximum (end-on view, panel (c)), the behaviour of $z_0$ is still well approximated by a gaussian, although our fit slightly under predicts the value of the scaleheight.

Along some cuts at \emph{x} values intermediate between the centre and the peanut maximum, the gaussian approximation fails to represent the behaviour of the scaleheight with \emph{y}. In fact the behaviour of the scaleheight in the presence of a B/P bulge is quite complex, and cannot be grasped entirely by a simple analytic function. However, as it turns out, the fitted function shown in Fig.~\ref{fig:fits} underestimates the value of $z_0$ at these points. This directly translates into an underestimation of the effect of the peanut in those regions. In summary, our fiducial model for the peanut height function shown in Fig.~\ref{fig:fits} will result into a conservative estimate of the effect of the real peanut present in the image we adopt as our starting point. Given the scope of this paper, which is to demonstrate the generic effect of a peanut bulge on a galaxy model, we find this approximation more than satisfactory.

\subsection{Boxy height function}
\label{sec:BHF}

The B/P bulge might at times have rather boxy isophotes. This could be due to projection effects, whereby the peanut is projected at such an angle that the isophotes appear boxy \citep{AthanassoulaMisiriotis2002}. However, boxy isophotes might be present even when the bar is seen side-on, i.e. they might be the real shape of the B/P bulge (see \citet{Patsisetal2002} for a discussion based on orbits). This tends to be the case for galaxies with weak bars, where instead of a strong x-shape or peanut forming, boxy isophotes are seen \citep{Athanassoula2006ph}.

To model a boxy bulge we use a height function which drops off as sech$^2$ with height from the \emph{z}=0 plane, 

\begin{equation}
F(x,y,z)=\frac{1}{2 z_0(x,y)}\mathrm{sech}^2\left(\frac{z}{z_0(x,y)}\right)  ,
\label{eq:boxytophat1}
\end{equation}

\noindent where the scaleheight is a top-hat function,

\begin{equation}
z_0(x,y) =
\begin{cases}
  z_0^{bulge} & |x| \leq x_{max} \,\&\, |y| \leq y_{max}\\
  z_0^{disc} & \text{otherwise}.
\end{cases}
,
\label{eq:boxytophat2}
\end{equation}

\noindent and where $z_0^{disc}$ gives the scaleheight of the disc and $z_0^{bulge}$ gives the scaleheight, or strength, of the boxy bulge.
This is quite a simplified model of the boxy bulge, with only two free parameters, its strength and length (which is set by \emph{L}=2\emph{$x_{max}$}). The thickness of the box, i.e. \emph{$y_{max}$}, is set by the width of the bar.

We create the fiducial boxy height function such that it best approximates the fiducial peanut height function, so such that we can examine whether the former can be used as an approximation for the latter, as there is one less parameter to model. The fiducial boxy height function therefore has a height equal to the height of the fiducial peanut model and its length is such that the boxy bulge finishes approximately where the peanut scaleheight is in between its maximum and minimum (see top right panel of Fig.~\ref{fig:resultsgtr116potforce}).

\section{Boxy/Peanut or no Boxy/Peanut?}
\label{sec:B/P-noB/P}

We wish to investigate whether accounting for the geometry of the B/P in the height function will significantly change the model of its host galaxy, and therefore whether we should include it in the modelling when it is present. Thus, in the next three subsections we investigate the effect B/Ps will have on the potential and forces (Section \ref{sec:models_pea_nopea}), on the periodic orbits (Section \ref{sec:pea-nopea-orbits}) and on the bar strength (Section \ref{sec:simbarstrength}) of the model. 
\subsection{B/P effect on potential and forces}
\label{sec:models_pea_nopea}

\begin{figure*}
\centering
\includegraphics[width=\textwidth]{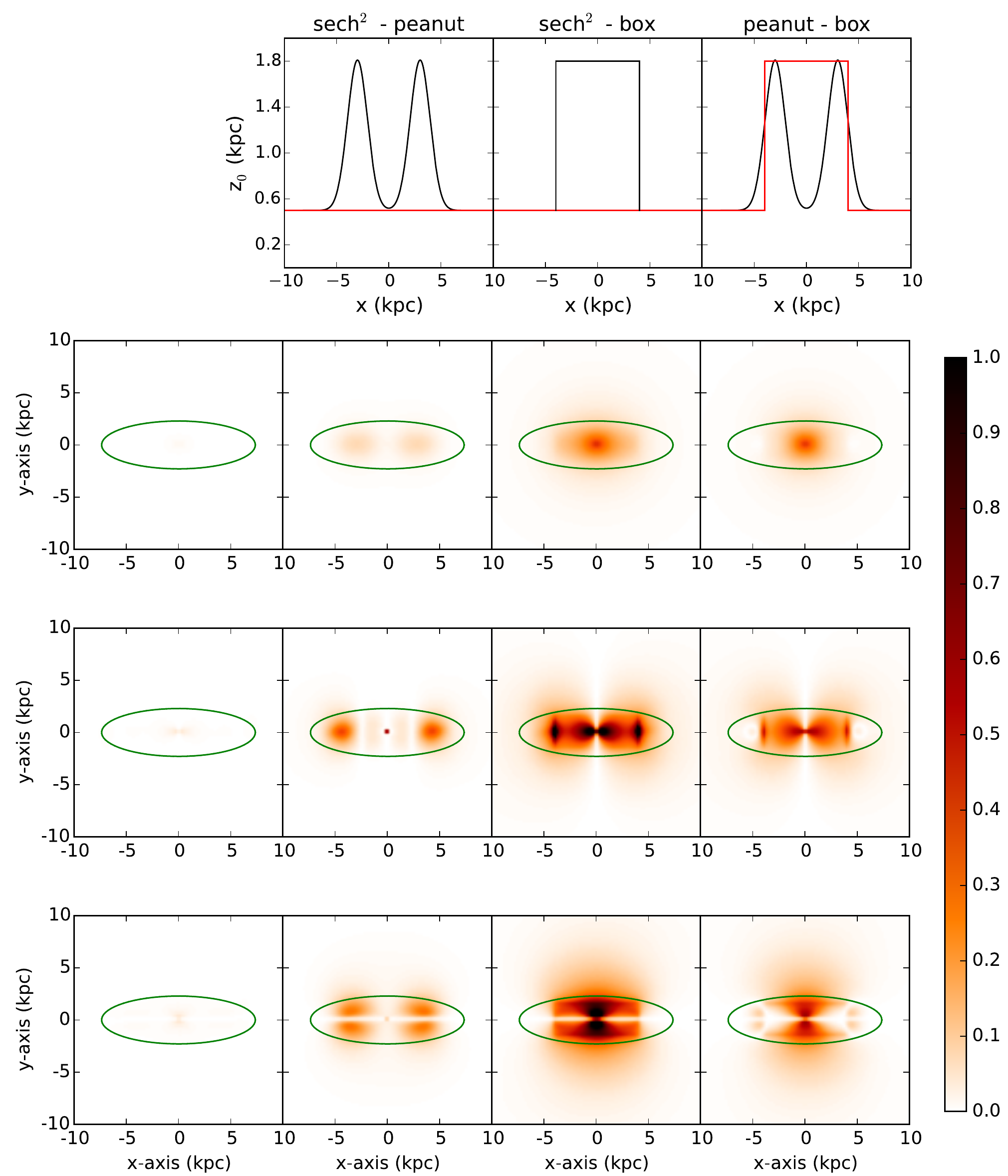}
\caption{\textbf{Errors from not taking into account the proper B/P geometry:} 
The top row gives the scaleheight (in red and black lines) along the bar major axis for the setups we are comparing in the plots. The second, third and fourth rows give the relative difference between the two setups being compared for the potential, $F_x$ and $F_y$, respectively (see the colourbar for values of the relative difference). The dark green line represents the ellipse fitted to the outer isophote of the bar.
\textit{First Column}: Difference between the sech and sech$^2$ setup.
\textit{Second Column}: Difference between the fiducial peanut height function and a sech$^2$ height function.
\textit{Third Column}: Difference between a boxy height function and a sech$^2$ height function.
\textit{Fourth Column}: Difference between a boxy height function and our fiducial peanut height function.
We see that not including a peanut or a boxy bulge where there is one will induce large errors in the potential and forces and also that a boxy height function is not a good approximation for a peanut height function. For details see the text (Section \ref{sec:models_pea_nopea}).} 
\label{fig:resultsgtr116potforce}
\end{figure*}

We calculate the potential and the forces for the density distribution given by the image described in Section \ref{sec:smoothing}, and the different height functions described in Section \ref{sec:section3}.
The results of this subsection are shown in Fig. \ref{fig:resultsgtr116potforce}. In the top row we plot the scaleheight along the bar major axis for the two setups we are comparing in the plots. We show two-dimensional plots of the relative difference for the potential in the second row, of the $x$ component of the force in the third row, and of the $y$ component of the force in the fourth row. The green line represents an ellipse fitted to the outer isophote of the bar. Each column gives one of the following comparisons (from left to right):

\begin{enumerate}
	
	\item \emph{Two flat height functions: sech and sech$^2$}
	
	We compare the models obtained by implementing two flat height functions, sech and sech$^2$, with equivalent scaleheights, in order to demonstrate that different flat height functions do not significantly affect the results. In the very centre, the difference for the potential is only around 1\% and for the forces it is 5\%, while for the rest of the grid the difference between the two setups is well below 1\% in all cases. If we decrease the value of the scaleheight the two height functions produce even more similar results. This happens because as the disc tends to become infinitesimally thin, the shape of the height function becomes less important.  Equivalently, if we increase the value of the scaleheight, and hence the thickness of the disc, then the difference in the results obtained with the two height functions increases.

We see therefore that the scaleheight, and not the vertical profile of the height function, is primarily responsible for creating differences in the models. \newline
	
	\item \emph{Flat and peanut height functions}
	
	We compare a flat height function and the height function of our fiducial peanut bulge, i.e. a peanut with parameters fitted to our simulated galaxy. The differences that arise from using these two setups are significant for the potential and forces, as can be seen in the second column of Fig. \ref{fig:resultsgtr116potforce}. This is especially true near and around the region of the maximum height of the peanut, and in general in and around the bar. The force can be different in the two cases by up to 40\%, which is not surprising since around the maximum of the peanut the scaleheight is more than three times the value of the scaleheight of the disc. This can be seen in the top row of the figure, which demonstrates how the scaleheight varies along $x$ for the two height functions. The larger scaleheight reduces the forces in the plane of the disc, due to a reduction of the density in the plane.

Therefore, we see that by not taking into account the geometry of the B/P bulge, we induce significant errors in the model, i.e. in the potential and its derivatives.\newline
	
	\item \emph{Flat and boxy height functions} 
	
	In the third column of Fig. \ref{fig:resultsgtr116potforce} we compare a flat height function and the height function of the fiducial boxy bulge. We see that a boxy height function will also induce large differences compared to the flat height function, and in fact in the central regions our fiducial boxy height function has an even larger effect than the peanut. Boxy bulges are usually associated to weaker bars in simulations and are therefore typically less strong than peanut bulges (although at early times boxy bulges can be as strong as peanut bulges - see Figs. 2 and 3 in \citet{AthanassoulaMisiriotis2002}). In observations, boxy bulges can appear as strong as peanut bulges (see for example \citet{Chung&Bureau2004}) although it is hard to distinguish whether these are truly boxy bulges, or simply peanut bulges seen at an angle. It is therefore reasonable to assume that a substantial amount of boxy bulges will be somewhat less strong than our fiducial boxy height function. However, for the sake of simplicity and to be able to compare our fiducial boxy height function with our fiducial peanut height function, we give the former the same strength as the latter. Therefore our fiducial boxy height function can be thought of as an upper limit for the effect of a boxy bulge on the model of its galaxy. \newline

	\item \emph{Peanut and boxy height functions} 
	
	We compare our fiducial peanut height function to our fiducial boxy height function in order to see to what extent the boxy height function approximates the peanut height function as it has one less free parameter than the peanut height function. These results can be seen in the fourth column of Fig. \ref{fig:resultsgtr116potforce}. For the potential and forces the match between the boxy height function and the peanut height function is quite poor, especially in the central region where the scaleheights of the two height functions are very different. Therefore the boxy height function is not a good approximation to a peanut bulge.\newline
	
\end{enumerate}

\subsection{B/P effect on periodic orbits}
\label{sec:pea-nopea-orbits}

\begin{figure*}
\centering
\subfigure[Typical orbits in bar region]{%
	\includegraphics[height=6.9cm]{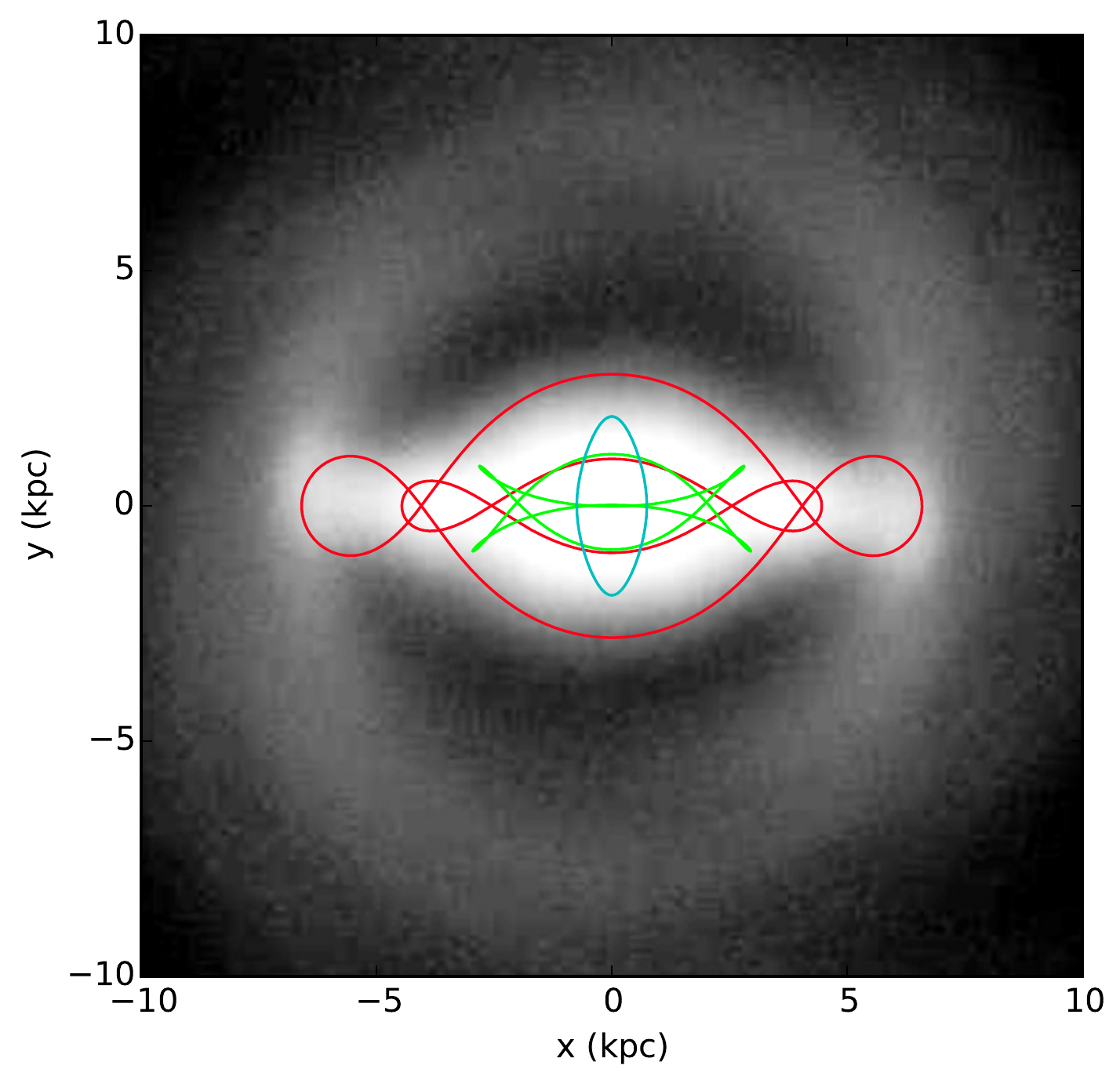}
	\label{fig:gtr116orbits}}
\quad
\subfigure[Characteristic Diagram]{%
	\includegraphics[height=6.9cm]{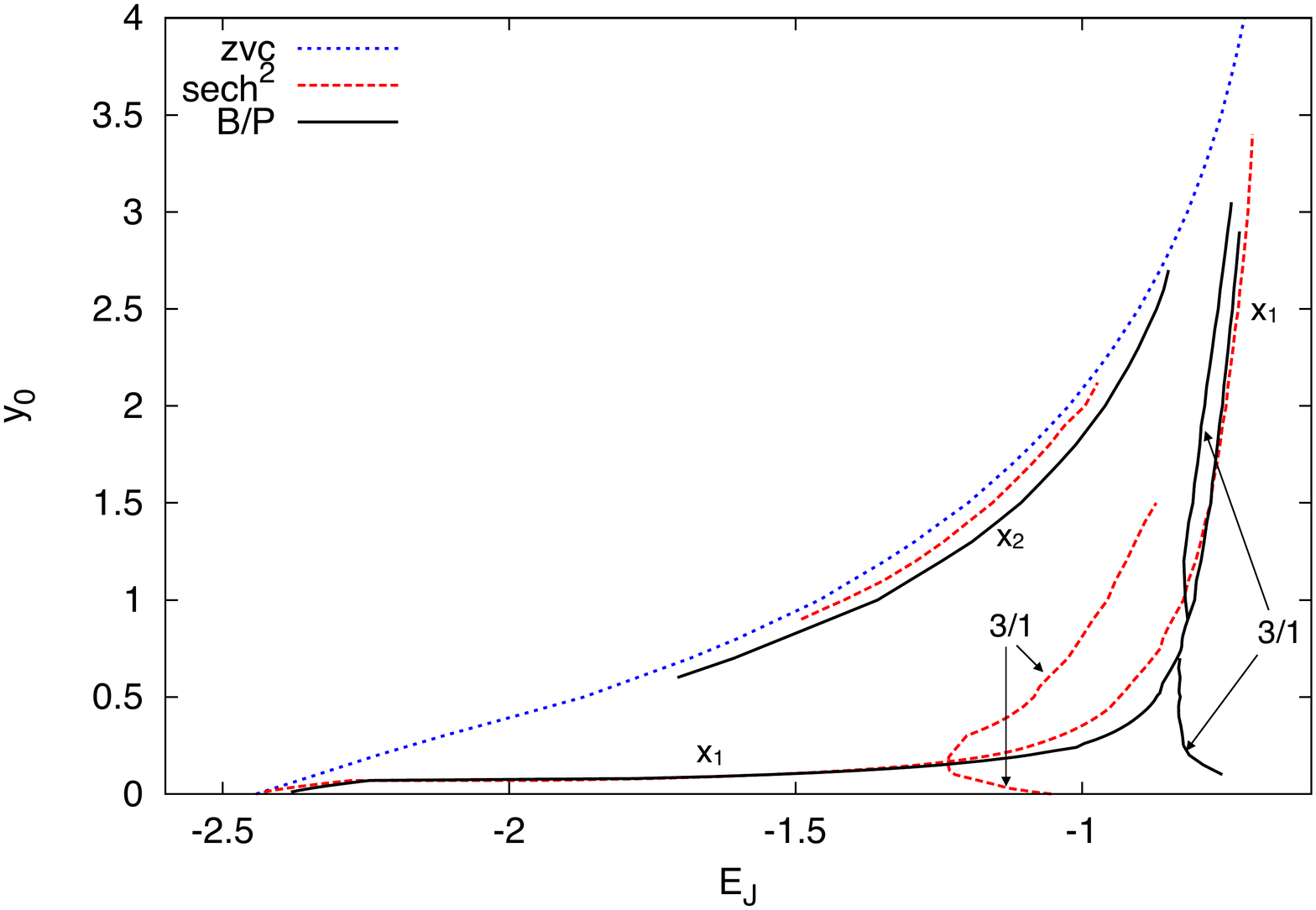}
	\label{fig:orbgtr116Ej}}
\quad
\caption{\emph{Left}: Some typical orbits in the bar region for the model with a sech$^2$ height function, over-plotted on the image of the simulated galaxy gtr116: In red the x$_1$ bar supporting orbits, in cyan an x$_2$ orbit perpendicular to the bar and in green the 3/1 orbits (asymmetric with respect to the $y$-axis). 
\emph{Right}: Characteristic diagram (intersection of each orbit with the $y$-axis as a function of the Jacobi energy) for the models created from the image of the simulated galaxy gtr116 and the two height functions. The solid black line gives the characteristic diagram for the model with the fiducial peanut bulge and the dashed red line the characteristic diagram for a model with a flat sech$^2$ height function. The dotted blue line shows the zero velocity curve (ZVC) for the sech$^2$ model (the ZVC of the two models are very similar).} 
\label{fig:orbsgtr116peanopea}
\end{figure*}

\begin{figure*}
\centering
\subfigure[Comparing x$_1$ orbits]{%
	\includegraphics[height=7.5cm]{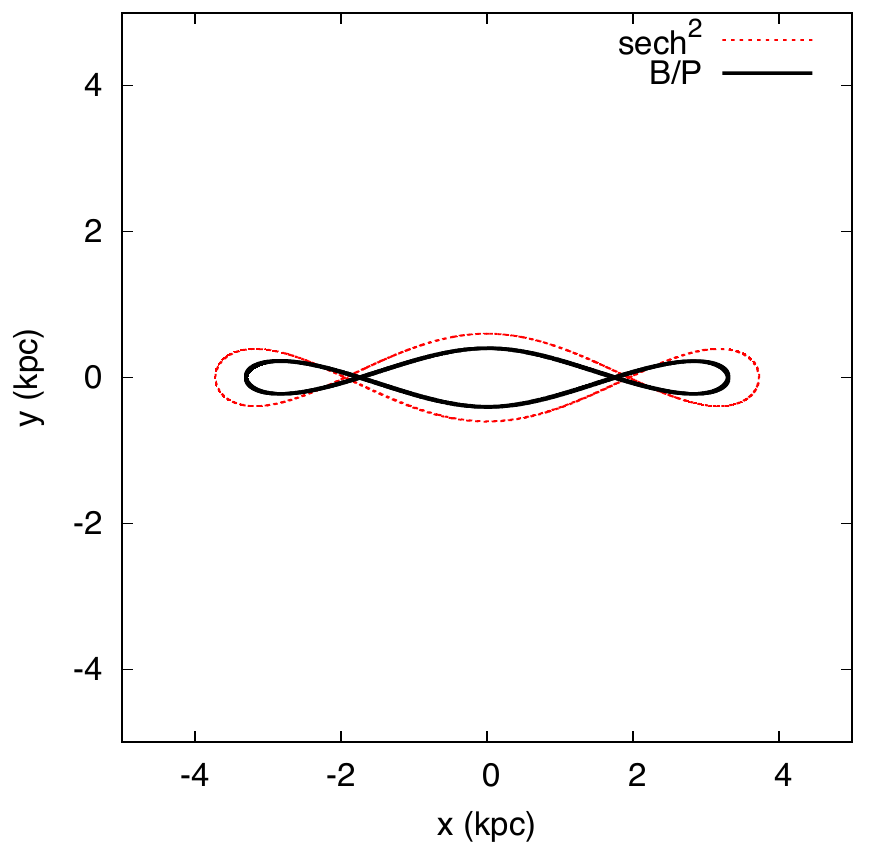}
	\label{fig:difx1orb}}
\quad
\subfigure[Comparing 3/1 orbits]{%
	\includegraphics[height=7.5cm]{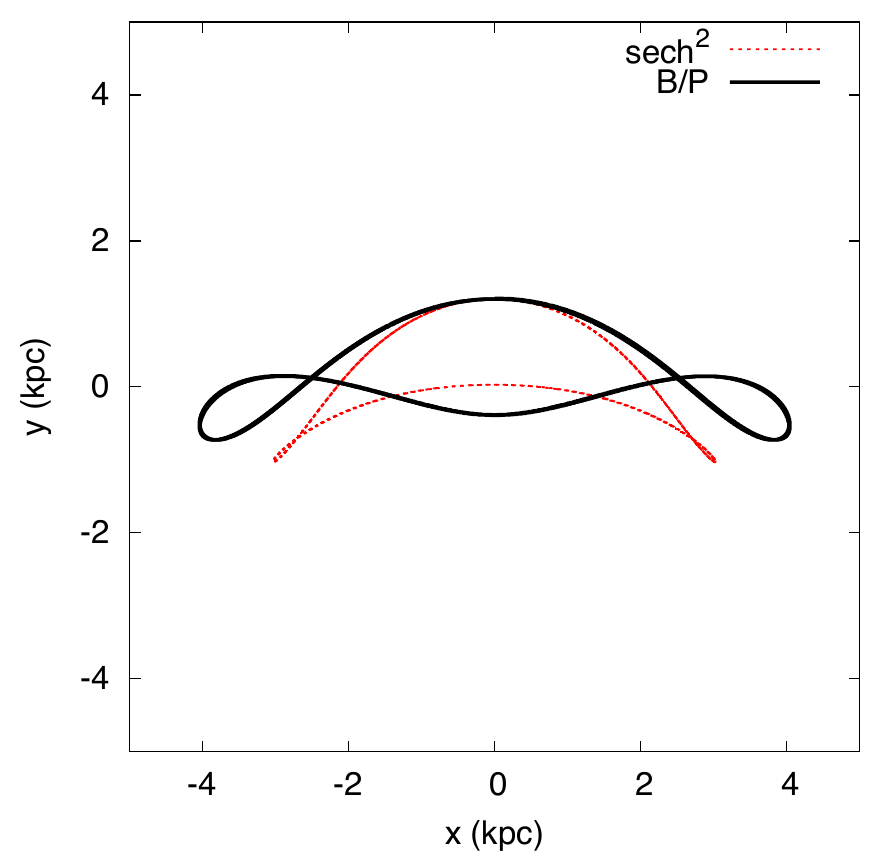}
	\label{fig:dif31orb}}
\quad
\caption{\emph{Left}: Two x$_1$ orbits with the same Jacobian energy ($E_J$=-0.9) calculated in the two potentials: with (solid black line) and without (dashed red line) a B/P bulge. We see that in the B/P model the x$_1$ orbit's length and height are reduced (its extent along the $x$-axis is reduced by $\sim$12\%  and along the $y$-axis by $\sim$46\% -measured respectively at $y$=0 and $x$=0). \emph{Right}: Two 3/1 periodic orbits for $y_0$=1.2 in the two potentials (colours as before). Again the orbits differ significantly.} 
\label{fig:orbsgtr116peanopea}
\end{figure*}

In this section we examine how some of the most important families of periodic orbits will be influenced by taking into account the geometry of a B/P. To do this we study two models: the model with the fiducial peanut bulge height function, and the model with the sech$^2$ height function {\it{without}} a B/P bulge. We set the pattern speed of both models to be such that corotation occurs just outside the bar radius, within the range $1.4 \textgreater R_{CR}/R_{bar} \textgreater 1$, where $R_{CR}$ and $R_{bar}$ are the corotation and bar radius respectively (e.g. \citealt{Athanassoula1992b}). The orbits are calculated in a frame of reference co-rotating with the pattern speed of the bar.

In Fig.~\ref{fig:gtr116orbits} we show a few typical orbits in the bar region for the potentials we are examining, overplotted on the image of our simulated galaxy, gtr116, shown face-on. The three most important families of orbits in the bar region are shown, i.e. the x$_1$ (red lines, extended along the bar major axis), x$_2$ (cyan lines, perpendicular to the bar major axis) and $3/1$ (green lines, asymmetric with respect to the $y$-axis) families, which are stable along most of their extent.

In Fig.~\ref{fig:orbgtr116Ej} we plot the characteristic diagram of periodic orbits, for the two cases with and without a peanut. 
The characteristic diagram gives the value at which the orbit intersects the $y$-axis ($y_0$) as a function of its Jacobian energy ($E_J$, i.e. energy in the rotating frame of reference; \cite{BT2008}). The Jacobian energy is in arbitrary units, since, as already mentioned, the mass is also in arbitrary units. 
We see that the characteristic diagram of the two models differs significantly. The most noticeable effect due to the presence of a B/P is the change in the bifurcation loci of the upper and lower branch of the 3/1 family. This indicates that taking into account the geometry of a B/P in the model changes the location of the 3:1 resonance, and therefore the 3/1 family of periodic orbits appears at higher energies. Thus orbits of the 3/1 family will differ in the two cases, as can be seen in Fig. \ref{fig:dif31orb}, where we plot two 3/1 orbits in the two models for the same cut along the $y$-axis.
The extent of the 3/1 family of orbits is also significantly increased for the case with a B/P bulge, surpassing the extent of the x$_1$ family of periodic orbits, which is in fact shorter compared to the x$_1$ family in the model without a B/P bulge.

The x$_1$ family also suffers changes, in the $E_J$ region between -1.1 and -0.8. In this area of the diagram, the maximum extent of the orbits along the $x$-axis reaches the region where the effect of the B/P is maximum; therefore, for these energies the orbits of the two models differ. In Fig. \ref{fig:difx1orb} we show an $x_1$ orbit of the same energy ($E_J$=-0.9) plotted in the two models. When a B/P is present, the maximum extent of the orbit along the $x$-axis is reduced by 12\%, while its maximum extent along the $y$-axis is reduced by 46\% (measured at $x$=0).

For the x$_2$ family, the highest (lowest) value of $y_0$ increases (decreases) for the model with a B/P bulge (see Fig.~\ref{fig:orbgtr116Ej}), i.e. the entire extent of the x$_2$ family is increased by about 43\% . As the extent of the x$_2$ family is related to the distance between the two Inner Lindblad Resonances (ILRs; \citealt{Athanassoula1992a}), the distance between these two resonances will therefore also increase. This increase is due to a weakening of the non-axisymmetric perturbation: when the geometry of B/Ps is taken into account in the model, the scaleheight of the galaxy is increased, where the B/P is maximum, and therefore -- since the amount of mass is the same-- the volume density in the plane of the galaxy is decreased. This leads to a decrease in the radial and tangential forces in such a way that the non-axisymmetric perturbation is diminished, thus changing the distance between the two ILRs. This is in accordance with results from both \cite{ContopoulosGrosbol1989} and \cite{Athanassoula1992a}, the latter of which showed that there are a number of model parameters which can affect this distance. In particular, in Figures 6 \& 7 of  \cite{Athanassoula1992a}, we see that the distance between the ILRs can increase due to a decrease in the bar mass or the pattern speed, or due to an increase in the central mass concentration of the galaxy or the axial ratio of the bar. 

The differences between the orbital families of the two models will have effects on their stellar, as well as their gaseous dynamics.
The extent of the x$_2$ orbits plays a crucial role on the shape of the shock loci in the gas  \citep{Athanassoula1992a,Athanassoula1992b}, and therefore the shape of the gas shocks in the two models should differ significantly; conversely, the shape and strength of the shocks influence the amount of gas inflow towards the centre of the galaxy, and it is therefore likely that there will be a measurable difference in the amount of gas inflow in models with and without a B/P bulge.
This is further supported by the results in Section \ref{sec:simbarstrength}, where we show that B/P bulges reduce the strength of the bar; we plan to address in upcoming work the extent to which the gas flows will be affected.
\subsection{B/P effect on bar strength}
\label{sec:simbarstrength}

\begin{figure*}
\centering
\includegraphics[width=0.65\textwidth]{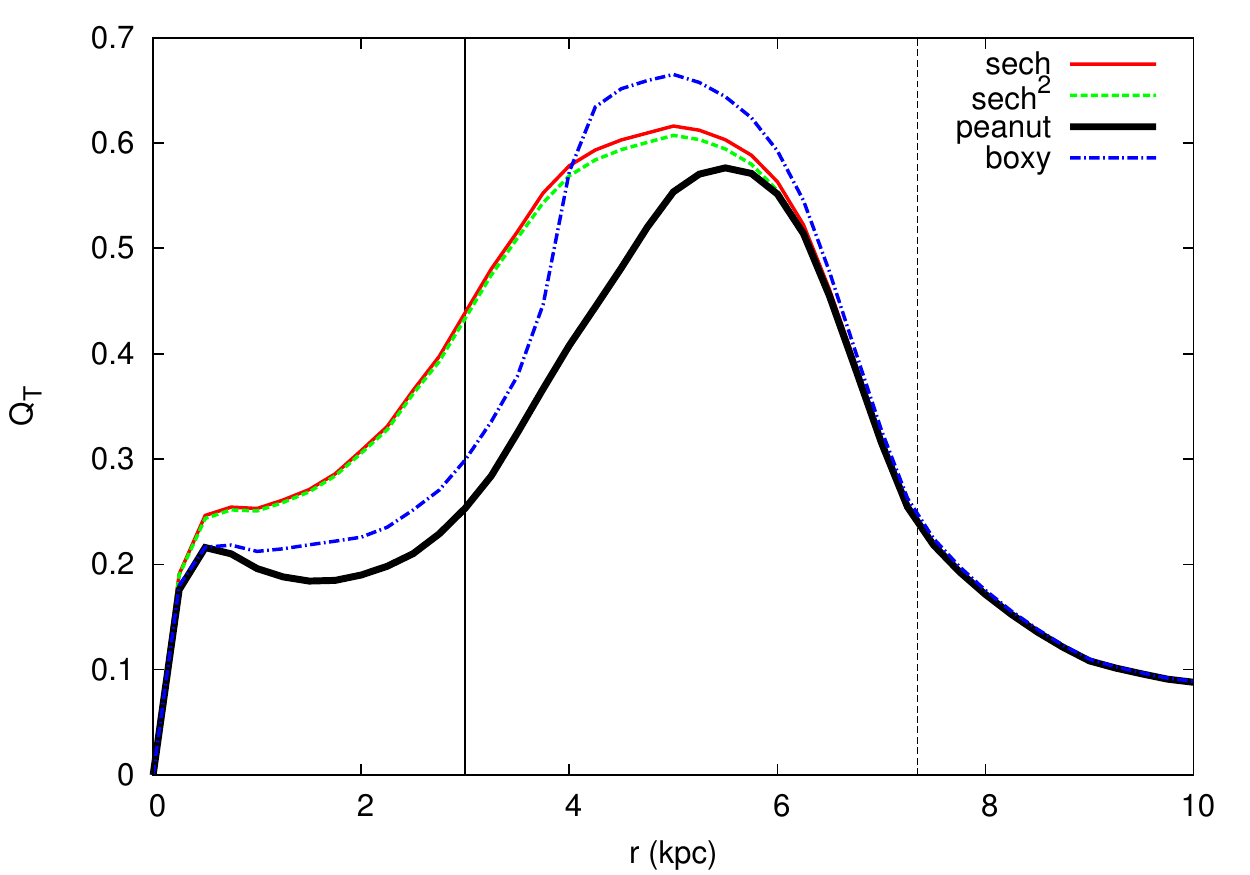}
\caption{Strength of non-axisymmetric forcings ($Q_T$) as a function of radius, for models with different height functions: sech (solid red line), sech$^2$ (dashed green line), the fiducial peanut setup (thick black solid line) and the fiducial boxy setup (dash-dotted blue line). The vertical solid black line indicates the radius at which the scaleheight of the fiducial peanut is maximum and the vertical dashed black line indicates the end of the bar.}
\label{fig:Qt_pea_nopea}
\end{figure*}

We study the effect of the B/P bulge on one of the measures of bar strength, which involves calculating the non-axisymmetric forcings on the disc due to the bar, i.e. the bar-induced torque \citep{CombesSanders1981, ButaBlock2001}. The magnitude of this non-axisymmetric perturbation is given by

\begin{equation}
Q_T(r)=\frac{F^{max}_{\mathrm{T}}(r)}{\langle F_{\mathrm{R}}(r) \rangle} ,
\label{eq:qt}
\end{equation}

\noindent where F$_T$ is the tangential force $F_T(r)$=(1/r) ($\partial \Phi$/$\partial \phi$), and $\langle F_R(r) \rangle$ is the average over azimuth of the radial force $F_R(r)$=$\partial \Phi$/$\partial r$. The forces are calculated directly from the image, as described in Section \ref{sec:section2}.
In order to obtain a single measure of the bar strength for a galaxy, the quantity $Q_b$, which is the maximum of $Q_T$ in the bar region, is commonly defined as the bar strength.
In what follows we investigate the effect that a B/P height function will have on both $Q_b$ and $Q_T$. 

The results of this study are discussed in paragraphs (\emph{i}), (\emph{ii}) and (\emph{iii}), where we examine models with the flat, the fiducial peanut and the fiducial boxy height function respectively. In Table~\ref{table:bp-nobp_strength} we show the maximum and average relative errors of $Q_T$, denoted MAX(Error $Q_T$) and $\langle$Error $Q_T$$\rangle$ respectively, as well as the relative error of $Q_b$, when comparing two models with different height functions. The average relative error of $Q_T$ over radius is calculated according to:

\begin{equation}
\langle\mathrm{Error} \;Q_{T}\rangle=\frac{1}{n} \times \sum_{i=1}^n (\mathrm{abs}(\frac{Q_{T_{1}}(r_i)-Q_{T_{2}}(r_i)}{Q_{T_{1}}(r_i)}) \times 100) .
\end{equation}

Plots of these results can also be seen in Fig.~\ref{fig:Qt_pea_nopea}, where it is worth noting that for the flat and peanut height functions, the strength of the bar is $Q_b$$\approx$0.55-0.6. According to \citet{ButaBlock2001}, this represents a strong bar case (between bar class 5 and 6), which corresponds to approximately 20\% of their sample of SB galaxies. We have already shown in Fig.~\ref{fig:IC4290} the striking morphological similarity between IC 4290 and our galaxy, and we note here that IC 4290 is also classified by \citet{ButaBlock2001} as a class 6 barred galaxy, with $Q_b$=0.56. Therefore the results presented in this section, as well as in previous and subsequent sections, correspond straightforwardly and quantitatively to strongly barred galaxies. However, even weakly barred galaxies will have B/P bulges, albeit weaker ones, and therefore the results will also apply to these galaxies although to a lesser extent. We intend to carry out a full statistical study of the effects of different strength B/P bulges on the models of their host galaxy, together with a full comparison to observations, elsewhere.

\begin {table}
\caption {Errors in bar strength}
\begin{center}
    \begin{tabular}{ r | r | r | r }
    \hline   
	Comparison & $\langle$Error $Q_T$$\rangle$ & MAX(Error $Q_T$) & $Q_b$ \\ \hline
    sech - sech$^2$ & 1\% & 1.6\% & 1.5\% \\ 
    peanut - no peanut & 27\% & 74\% & 4\% \\ 
    peanut - boxy & 14\% & 42\% & 16\%  \\ \hline
    \end{tabular}
\end{center}
We show the average and maximum of the relative errors of $Q_T$, as well as the relative error of $Q_b$, for three different comparisons of the setups.
\label{table:bp-nobp_strength}
\end{table}

\begin{enumerate}
	\item \emph{A model with a sech and sech$^2$ height function}

In previous work by \citet{HeikiEija2002} the effect of position-independent height functions and height functions which only vary as a function of radius was examined, and these were found not to change $Q_b$ in a significant way. For realistic height functions such as the exponential, sech, or sech$^2$ models, they found that $Q_b$ was affected by less than 5\%, which is consistent with our own results. This is confirmed in Fig. \ref{fig:Qt_pea_nopea} and Table \ref{table:bp-nobp_strength}, where we see that for an equivalent scaleheight, the height functions of sech and sech$^2$ will produce very similar bar strengths, which will tend to become even more similar the thinner the disc.\newline

	\item \emph{A model with the fiducial peanut height function}
	
We plot $Q_T$ for our fiducial B/P height function. In the region around the scaleheight maxima (at r=3), $Q_T$ is significantly flatter than the model without a peanut, due to the reduction of the strength of the tangential and radial forces. The value of $Q_b$ will not be significantly different from the case with the flat height function, due to the fact that the maxima of the peanut and the maximum of $Q_T$ are at a relatively large distance from each other. The torque induced by the bar in the two cases however is significantly different, as can be seen both in the plot and in the second row of Table \ref{table:bp-nobp_strength}. By using the $Q_b$ method of measuring bar strength, the bars will be judged as having the same strength (class 6), and hence the same effect on the disc, even though the forces in the plane of the galaxy are significantly reduced in the presence of a peanut. Therefore, in Section \ref{sec:Qt_pea-pea} we introduce another measure of bar strength, which can capture the reduction in bar strength when a B/P bulge is present.\newline

	\item \emph{A model with the fiducial boxy height function}
	
We also plot $Q_T$ for our fiducial boxy height function. We see again that where the boxy bulge is maximum, $Q_T$ is flattened due to the decrease in the strength of the bar forces in the plane. We also see that where the top-hat boxy function ends, $Q_T$ exceeds the values of the sech and sech$^2$ curves. This is due to the effect of the boxy bulge on the tangential and radial forces, with the former increasing just outside the boxy bulge while the latter is decreased in the whole disc due to the overall decrease in mass-density in the plane of the galaxy. The combination of these two effects results in the torque becoming large in the region just outside the boxy bulge. As a consequence, $Q_b$ is overestimated by 16\% compared to the fiducial peanut case (third row of Table \ref{table:bp-nobp_strength}). We investigate the boxy height function as an alternative to using the peanut height function - since there is one parameter less to model - and conclude that even if one is merely interested in $Q_T$, a simple boxy height function is not a good approximation to a peanut function.

\end{enumerate}


\section{Errors due to Boxy/Peanut Modelling}
\label{sec:peanuterrors}

In this section we investigate how much error will be induced if we include a B/P bulge in the model, but with the main peanut parameters differing from that of our fiducial model. This type of error is induced due to observational uncertainties, as it is not trivial to observationally obtain the correct parameters for the B/P bulge we want to model. This is due to the physics of the problem, not the numerical part of the calculation (as is the error referred to in Section \ref{sec:relerrorstests} which can be made arbitrarily small) and is therefore practically an unavoidable source of uncertainty. Nevertheless, as we will discuss below, there do exist empirical and theoretical arguments which can constrain the parameter space of a B/P bulge.

The height function we have chosen for our fiducial B/P bulge, the sum of two two-dimensional gaussians, has three degrees of freedom. Thus inaccuracies in the modelling of the B/P bulge can also be introduced in three ways: by estimating wrongly the height of the gaussians (which corresponds to a change in peanut strength), or the distance between the maxima of the gaussians (which corresponds to a change in the peanut length), or the widths of the gaussians (which corresponds to a change in peanut `width', i.e. how peaked or thin the peanut is at its maximum).

\subsection{Potential and forces}
\label{sec:peanuterrors_potfor}

In this subsection we investigate how much error is introduced in the potential and forces by incorrectly modelling the B/P bulge.

\subsubsection{Peanut strength uncertainties}

The maximum value of the scaleheight of the peanut, also called the peanut strength, is a value which is not trivial to find observationally. Numerical studies have shown that the strength of the peanut correlates with the bar strength (Athanassoula 2006). However this relation has a considerable scatter, and can merely give an approximate estimate. \citet{Debattistaetal2005} showed that for face-on, or nearly face-on $N$-body simulated galaxies, an indicator of the presence and strength of a B/P bulge is the fourth order Gauss-Hermite moment of the line-of-sight velocity dispersion, $h_4$. However this relation has not been quantified in such a way which would allow a direct measurement of peanut strength from $h_4$. 
Studies of orbital structure \citep{Patsisetal2002,Skokosetal2002} have also suggested specific families of periodic orbits which are responsible for giving the B/P bulge its height, but no direct measurement of the strength of the B/P bulge is available from orbital structure either.

In order to measure the error due to peanut strength uncertainties, we use a grid of models with varying peanut strength and compare them to our fiducial peanut setup. These results can be seen in Table \ref{table:error_pea_strength} below. In this and in all subsequent tables, the term `Average Error', indicates the average error within the outer isophote of the bar and `Maximum Error' corresponds to the maximum error found in the grid, excluding the central-most point.
We see that an over- or under-estimation of the error by the same amount will produce similar errors in the model. In all the cases studied, the error induced by an incorrect peanut strength is always less than that induced by not modelling a B/P bulge at all.

\begin {table}
\caption {Percentage Error due to Peanut Strength Uncertainty}
\begin{center}
    \begin{tabular}{ r | r r r | r r r  }
    \hline
	\multirow{2}{*}{Peanut Strength} & \multicolumn{3}{c}{Average Error} & \multicolumn{3}{c}{Maximum Error} \\ 
	\cline{2-7}
	& $\Phi$ & $F_x$ & $F_y$ & $\Phi$ & $F_x$ & $F_y$ \\ \hline
    +50\% & 1.3\% & 2\% & 3\% & 3\% & 10\% & 6\%\\ 
    +25\% & 0.7\% & 1\% & 2\% & 1.5\% & 5\% & 3\% \\ 
    -25\% & 0.7\% & 1\% & 2\% & 1.5\% & 6\% & 4\% \\
    -50\% & 1.4\% & 3\% & 4\% & 3\% & 14\% & 10\% \\ 
    no peanut & 3\% & 8\% & 10\% & 7\% & 37\% & 28\%\\ \hline
    \end{tabular}
\end{center}
Average and maximum errors of the potential and forces for setups with different peanut strength error, within the area enclosed by the outer isophote of the bar. The last row gives the error induced by not including a B/P bulge in the model at all.
\label{table:error_pea_strength}
\end{table}


\subsubsection{Peanut width uncertainties}

In Table \ref{table:error_pea_width} we show the errors for a grid of models with different peanut width errors.
For all cases considered, the error induced due to a miscalculation of the peanut width is less than that induced by not modelling a peanut at all, apart from the maximum error induced in the potential when the peanut width is 50\% larger than in the fiducial scenario. This is due to a sharp increase in scaleheight in the central region for large peanut widths (see solid red line in Fig.~\ref{fig:z0_peawidth}), which is where the potential is most affected. This error however is confined only to the potential and to the central most grid points, and should not have significant effects on orbital calculations in most of the galaxy.

\begin {table}
\caption {Percentage Error due to Peanut Width Uncertainty}
\begin{center}
    \begin{tabular}{ r | r r r | r r r  }
    \hline
	\multirow{2}{*}{Peanut Width} & \multicolumn{3}{c}{Average Error} & \multicolumn{3}{c}{Maximum Error} \\ 
	\cline{2-7}
	& $\Phi$ & $F_x$ & $F_y$ & $\Phi$ & $F_x$ & $F_y$ \\ \hline
    +50\% & 3\% & 5\% & 6\% & 15\% & 26\% & 27\% \\ 
    +25\% & 1\% & 2\% & 3\% & 4\% & 10\% & 12\% \\  
    -25\% & 1\% & 2\% & 3\% & 3\% & 11\% & 6\% \\
    -50\% & 2\% & 5\% & 6\%  & 5\% & 11\% & 15\% \\ 
	no peanut & 3\% & 8\% & 10\% & 7\% & 37\% & 28\%\\ \hline
    \end{tabular}
\end{center}
Average and maximum errors of the potential and forces for setups with different peanut width errors within the area enclosed by the outer isophote of the bar. The last row gives the error that would be present if we do not model a B/P bulge at all.
\label{table:error_pea_width}
\end{table}


\subsubsection{Peanut length uncertainties}
 
Of the three parameters - length, strength and width - length has the least uncertainty, due to a method proposed by \citet{Athanassoulaetal2014}. The method determines the length of B/Ps for face-on and moderately inclined galaxies, which uses the shape of the projected isophotes in the bar region.
They demonstrated that the barlens and the peanut are the same component and therefore that the size of the former can be used to estimate the length of the latter. For galaxies with larger inclinations the length can be estimated from other morphological features in the isophotes created by the B/P bulge \citep{AthanassoulaBeaton2006, ErwinDebattista2013}, while orbital structure studies confirm the aforementioned results and also give clues as to the length of the peanut \citep{Patsisetal2002,Patsisetal2003}. Due to all this, the uncertainties of the length estimates are rather small, certainly smaller than the corresponding ones for strength and width, which is why we use a smaller range of uncertainties for the peanut length.

We carry out comparisons for a grid of models with different peanut length errors and give the results in Table \ref{table:error_pea_length}. As expected, the more we change the length of the peanut away from the fiducial value, the larger the errors will be, although there is an asymmetry in the error induced with respect to over- and under-estimating the length; by underestimating the peanut length by a certain amount, we induce more error than by overestimating it by the same amount. By decreasing the length of the peanut we induce more error in the central regions of the galaxy, which is where the potential is most affected, due to the two gaussians overlapping in the centre and thus increasing the scaleheight (see dotted magenta line, Fig.~\ref{fig:z0_pealength}). This can be seen in Table \ref{table:error_pea_length}, for the case of -30\% peanut length where, for the potential, the maximum and average errors are larger than that of the +30\% peanut length case.

For all the cases considered, the average error induced is smaller or equal to that of not modelling the B/P bulge. Given that the length is a fairly well constrained quantity, large errors are not expected to be present due to the peanut length in the modelling of the B/P bulge. 

\begin {table}
\caption {Percentage Error due to Peanut Length Uncertainty}
\begin{center}
    \begin{tabular}{ r | r r r | r r r  }
    \hline
	\multirow{2}{*}{Peanut Length} & \multicolumn{3}{c}{Average Error} & \multicolumn{3}{c}{Maximum Error} \\ 
	\cline{2-7}
	& $\Phi$ & $F_x$ & $F_y$ & $\Phi$ & $F_x$ & $F_y$ \\ \hline
    +30\% & 1\% & 6\% & 5\% & 4\% & 24\% & 13\% \\ 
    +16\% & 0.9\% & 3\% & 3\% & 3\% & 12\% & 7\%  \\ 
    -16\% & 1\% & 4\% & 4\% & 5\% & 11\% & 10\% \\ 
    -30\% & 3\% & 8\% & 9\% & 13\% & 25\% & 28\%  \\ 
	no peanut & 3\% & 8\% & 10\% & 7\% & 37\% & 28\%\\ \hline
    \end{tabular}
\end{center}
Average and maximum errors of the potential and forces for setups with different peanut length errors within the area enclosed by the outer isophote of the bar. The last row gives the error induced by not modelling a B/P bulge at all.
\label{table:error_pea_length}
\end{table}


\subsubsection{Combinations of uncertainties}
\label{sec:comb_potfor}

It is likely that a combination of different kinds of error will contribute to the total error budget of a model of the B/P. It is not in the scope of this paper to explore the full parameter space of the possible error combinations, instead we choose a few cases in order to get a feel of the amount of error that can be induced. By `combination of error' we refer to a combination of all the different sources of error. By `+50\%' (`-50\%') we refer to a setup with peanut strength and width which are 50\% larger (smaller) than the fiducial value, and a peanut length which is 30\% larger (smaller) than the fiducial. We do not find it necessary to further increase the error in peanut length, since it is the best constrained quantity out of the three parameters. The `+25\%' (`-25\%') setup corresponds to one with peanut strength, width and length which is 25\% larger (smaller) than the fiducial value. In Table \ref{table:error_pea_allerrors} we see that all the combinations of uncertainties will introduce less error in the model than not including a B/P bulge at all. 

\begin {table}
\caption {Percentage Error due to Combination of Uncertainties}
\begin{center}
    \begin{tabular}{ r | r r r | r r r  }
    \hline
	\multirow{2}{*}{Combination} & \multicolumn{3}{c}{Average Error} & \multicolumn{3}{c}{Maximum Error} \\ 
	\cline{2-7}
	& $\Phi$ & $F_x$ & $F_y$ & $\Phi$ & $F_x$ & $F_y$ \\ \hline
    +50\% & 2\% & 3\% & 5\% & 6\% & 23\% & 22\% \\ 
    +25\% & 1\% & 4\% & 4\% & 6\% & 19\% & 18\% \\  
    -25\% & 1\% & 6\% & 6\% & 5\% & 29\% & 22\% \\ 
    -50\% & 2\% & 7\% & 8\%  & 6\% & 35\% & 26\% \\
	no peanut & 3\% & 8\% & 10\% & 7\% & 37\% & 28\%\\ \hline
    \end{tabular}
\end{center}
Average and maximum errors of the the potential and forces for setups with different combinations of errors within the area enclosed by the outer isophote of the bar. The last row gives the error induced by not modelling a B/P bulge at all.
\label{table:error_pea_allerrors}
\end{table}

\subsection{Periodic orbits}
\label{sec:pea-pea-orbits}

\begin{figure*}
\centering
\subfigure[]{%
	\includegraphics[height=6.5cm]{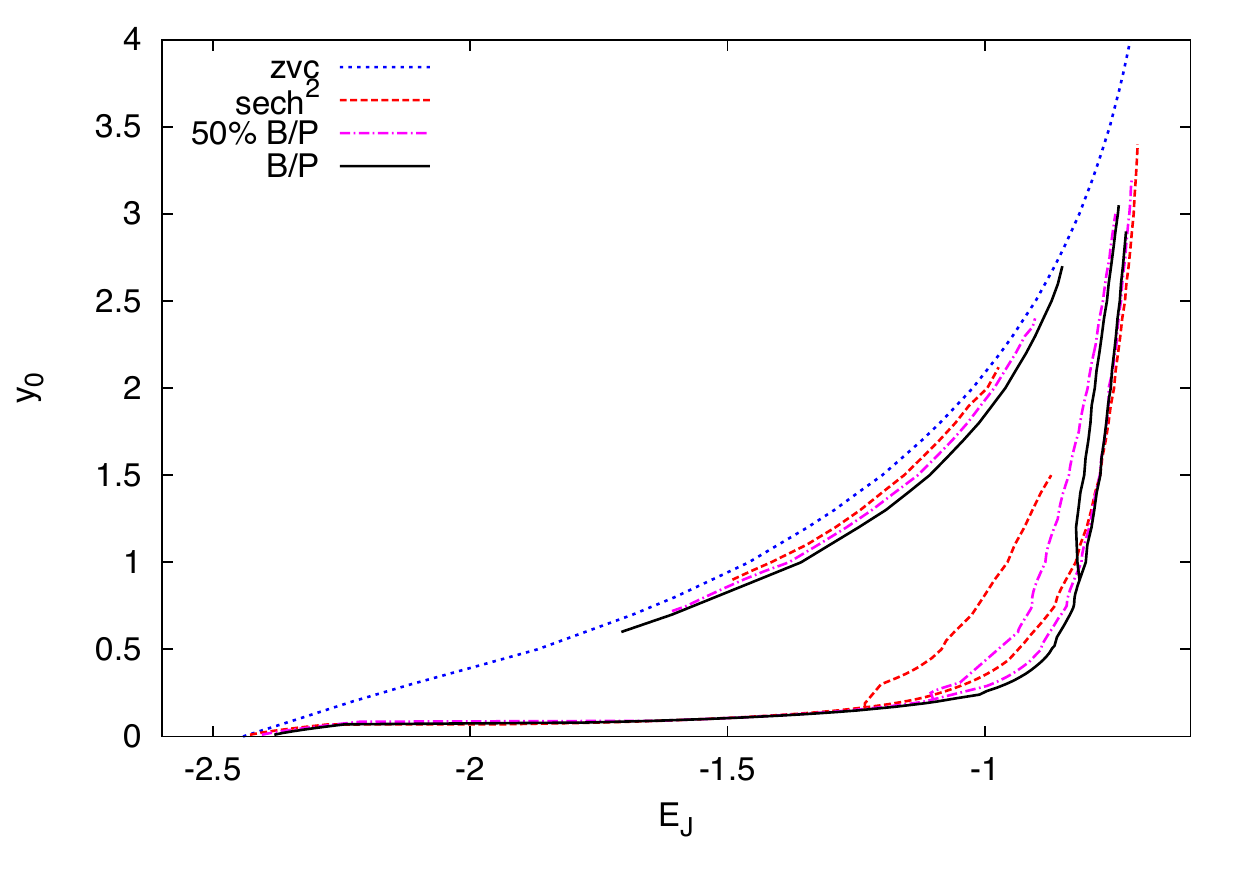}
	\label{fig:Ej_orb_peaerrors1}}
\quad
\subfigure[]{%
	\includegraphics[height=6.5cm]{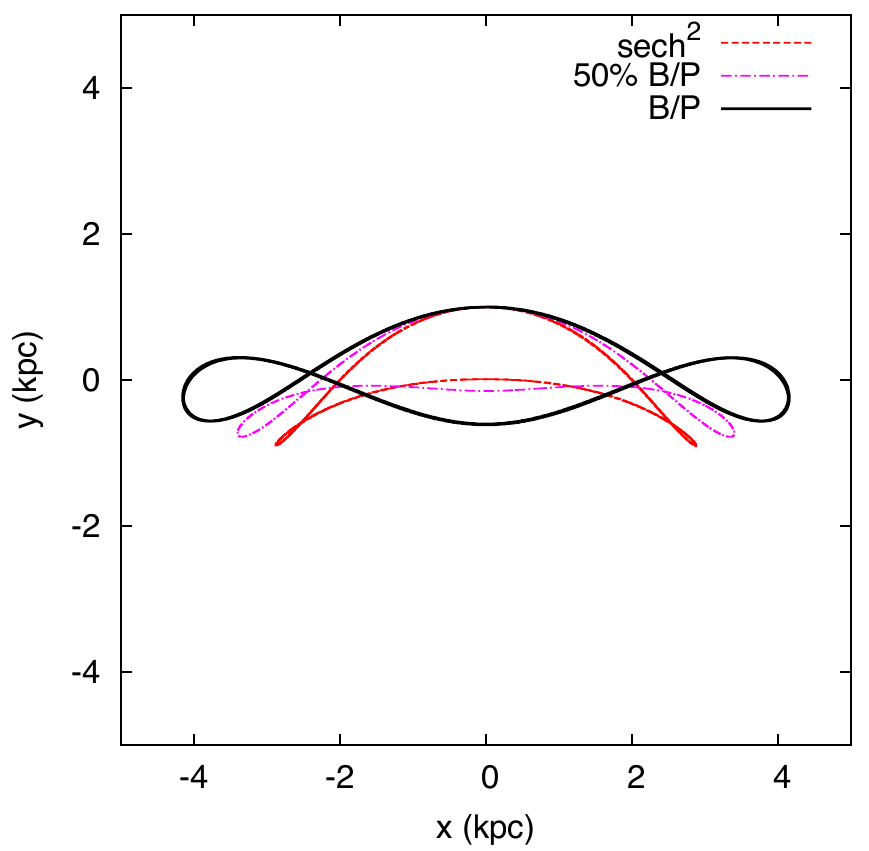}
	\label{fig:orbs_peaerrors}}
\quad
\caption{\emph{(a)} Characteristic diagram for models with different B/P setups. See Fig.~\ref{fig:orbsgtr116peanopea} and the text in Section \ref{sec:pea-nopea-orbits} for more details on the interpretation of the characteristic diagram. The dotted blue line gives the zero velocity curve; the characteristic diagram for the model with 0\%, 50\% and 100\% the fiducial peanut strength is given by the dashed red line, the magenta dashed-dotted line and the solid black line respectively.
\emph{(b)} The same $3/1$ orbit, which cuts the $y$-axis at $y$=1, in three models: without a B/P bulge (dashed red line), with 50\% peanut strength (magenta dashed-dotted line) and with 100\% peanut strength (black solid line).}
\label{fig:Ej_orb_peaerrors}
\end{figure*}

As shown in Section \ref{sec:pea-nopea-orbits}, the presence of a B/P bulge will affect the extent and shape of the different families of periodic orbits which make up the bar. In this section we qualitatively explore the errors introduced in the calculation of periodic orbits due to incorrect modelling of a B/P bulge. To do this we examine the characteristic diagram of the most relevant families of periodic orbits for three models with different peanut strengths (100\%, 50\% and 0\% of the fiducial strength), shown in Fig. \ref{fig:Ej_orb_peaerrors1}.

For the model with 50\% the fiducial strength the extent of the x$_2$ orbits is reduced by $\sim$19\%, while if we do not add a B/P at all, the extent of the x$_2$ family is reduced by $\sim$43\% (more than double the error for the 50\% peanut strength case). 

The bifurcation locus of the 3/1 family for the model with 50\% the peanut strength occurs about halfway between the locus of the models with and without a B/P bulge. Additionally, the extent of the 3/1 family in the characteristic diagram for this model is almost the same as for the model with the fiducial peanut, while the extent of the 3/1 family without a B/P is significantly shortened. In Fig. \ref{fig:orbs_peaerrors} we can see how the 3/1 orbits are affected by the incorrect modelling of the B/P bulge: for the same cut along the $y$-axis the 3/1 orbits are more elongated in the case with the fiducial peanut model, while they become less elongated and more concave with respect to the bar as the peanut strength is reduced. However, as expected, the orbits in the model with 50\% the peanut bulge better match the orbit of the fiducial B/P setup than the model without the B/P bulge.

As already noted in Section \ref{sec:pea-nopea-orbits}, the x$_1$ family of orbits is also affected by the presence of the B/P bulge, when the maximum extent of the orbits reach the region where the effect of the B/P is maximum, i.e. around ($x$, $y$)=($\pm3\,\mathrm{kpc}$, $0\,\mathrm{kpc}$). On the characteristic diagram this occurs in the region around ($E_J$, $y_0$)=(0.9, $0.5\,\mathrm{kpc}$). However even by underestimating the strength of the B/P by 50\%, the x$_1$ family is quite similar to the x$_1$ family of the fiducial B/P case.

We see that in general, the characteristic diagram of the model with 50\% the fiducial strength has features which are more towards the fiducial peanut model and therefore even with such a large error in peanut strength, the characteristic diagram of this model reproduces relatively well the characteristic diagram of the fiducial B/P model and certainly better than the model without a B/P bulge.
Similar results are found when considering errors in peanut width and length, and we therefore conclude that it is preferable to include a B/P in the model; the orbital structure of the model is significantly affected when a B/P bulge is present, and by adding a B/P, even with large errors in its parameters, the periodic orbits reproduce the correct structure more closely than when not including a B/P at all. 

\subsection{Bar strength}
\label{sec:Qt_pea-pea}

\begin{figure*}
\centering
\subfigure[]{%
	\includegraphics[width=0.45\textwidth]{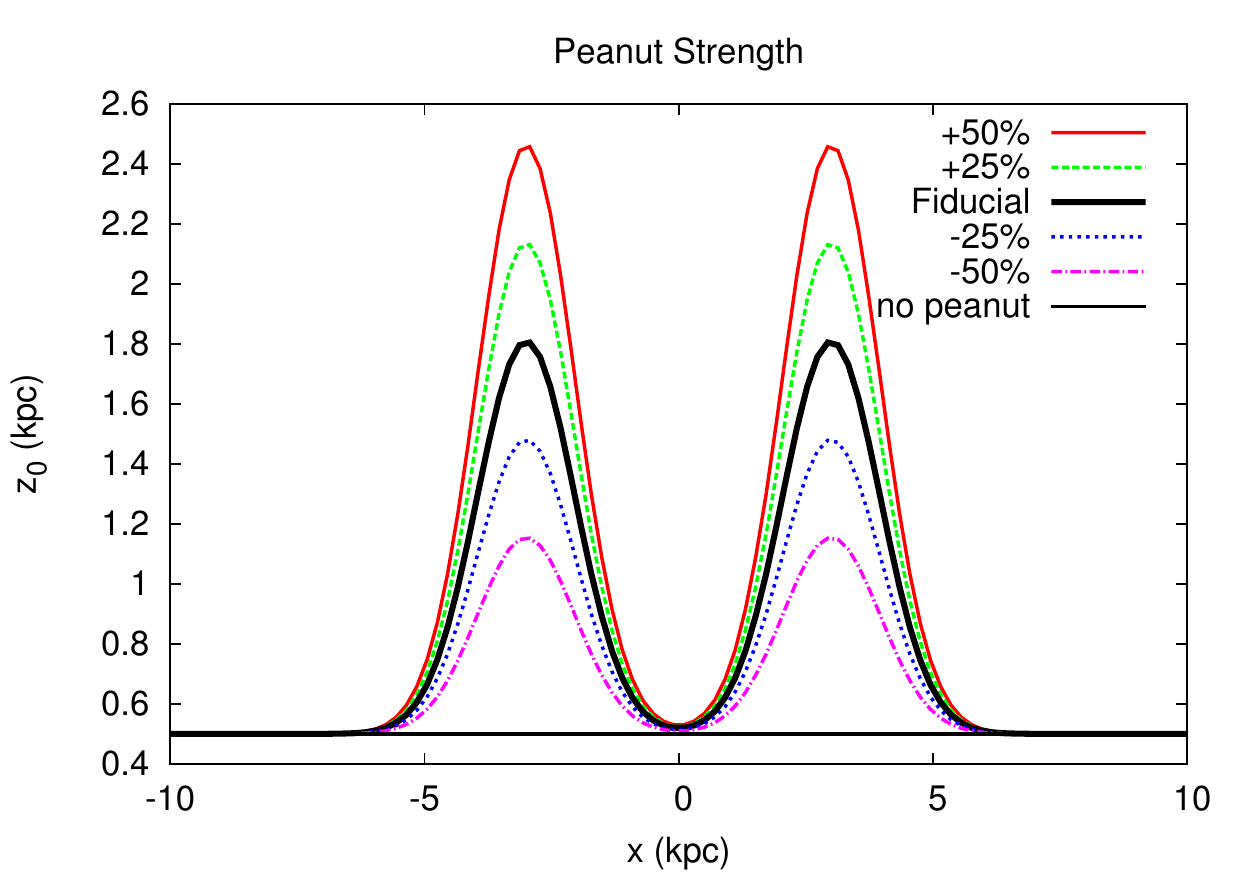}
	\label{fig:z0_peastrength}}
\quad
\subfigure[]{%
	\includegraphics[width=0.45\textwidth]{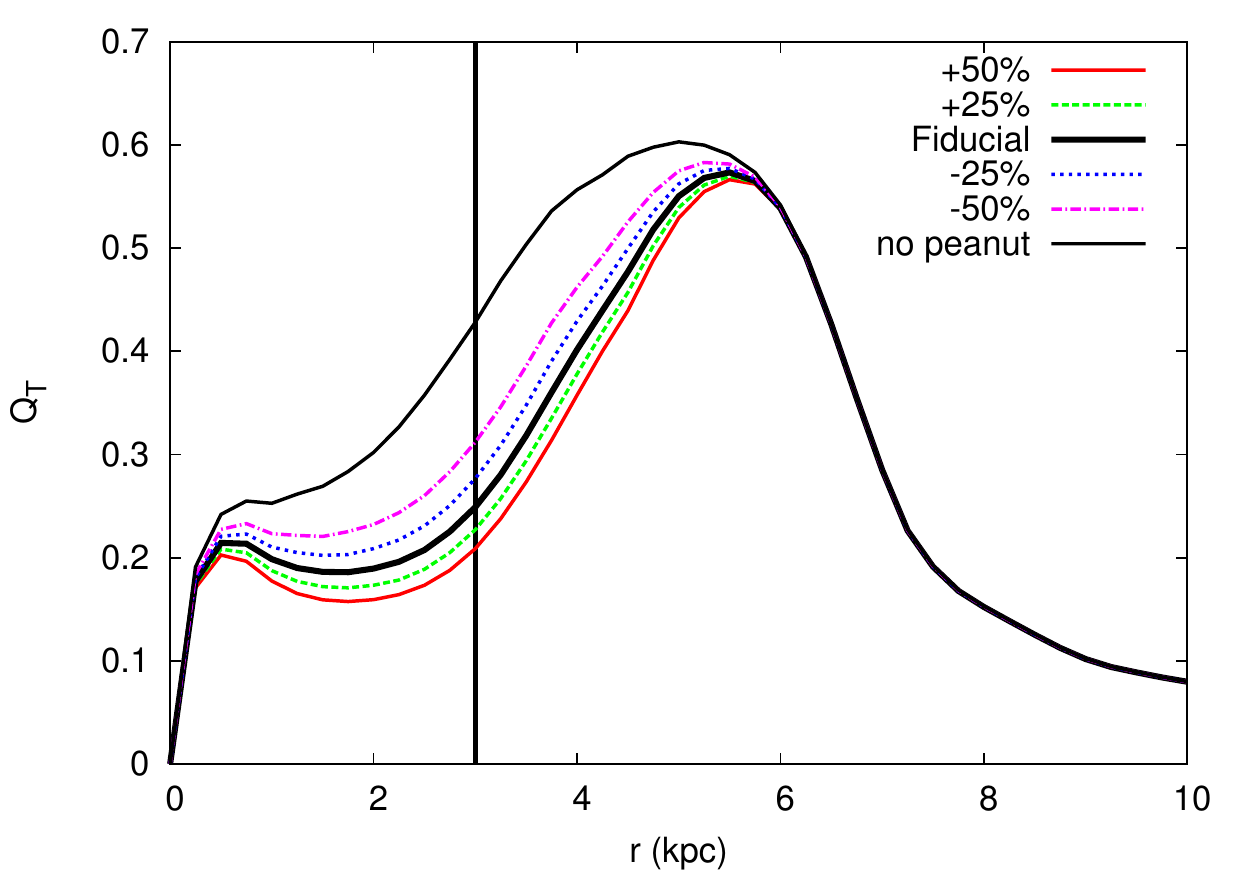}
	\label{fig:Qt_peastrength}}
\quad
\subfigure[]{%
	\includegraphics[width=0.45\textwidth]{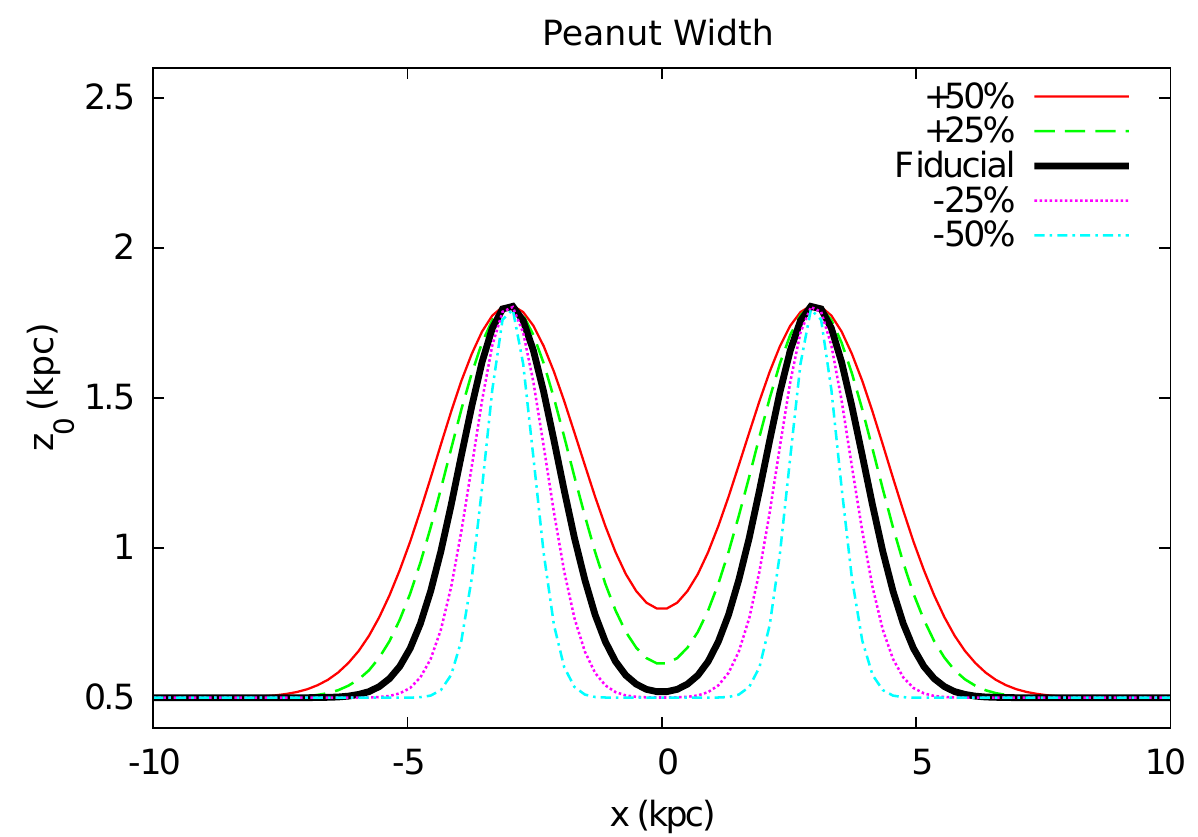}
	\label{fig:z0_peawidth}}
\quad
\subfigure[]{%
	\includegraphics[width=0.45\textwidth]{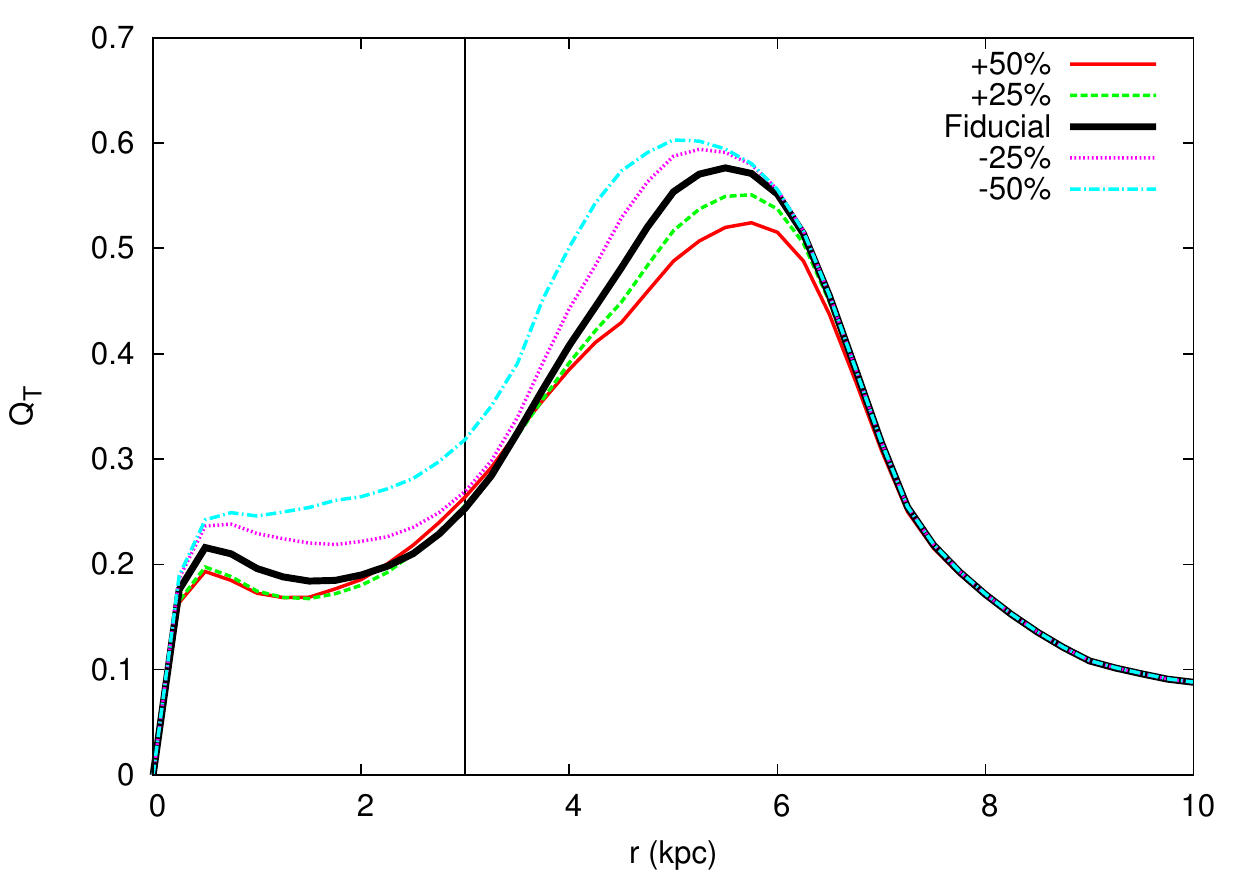}
	\label{fig:Qt_peawidth}}
\quad
\subfigure[]{%
	\includegraphics[width=0.45\textwidth]{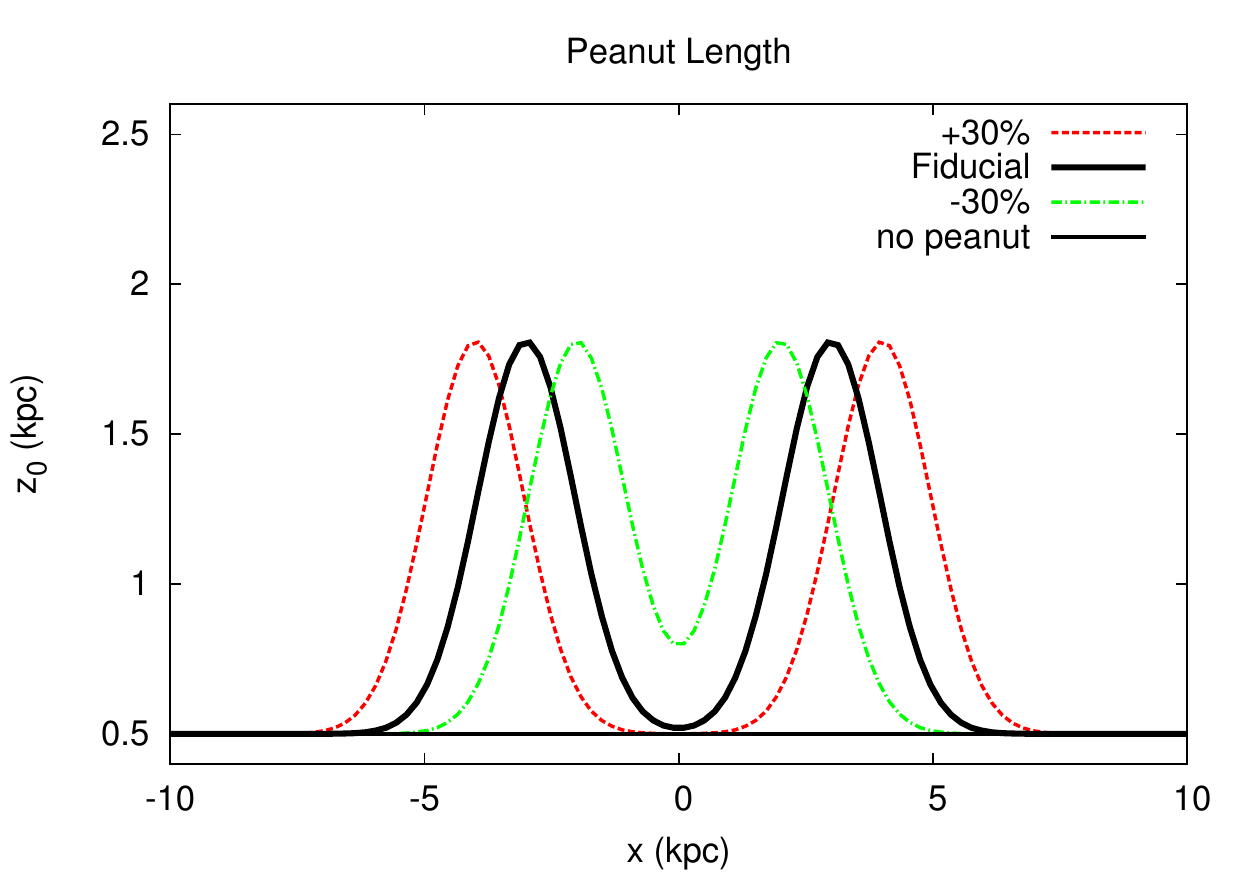}
	\label{fig:z0_pealength}}
\quad
\subfigure[]{%
	\includegraphics[width=0.45\textwidth]{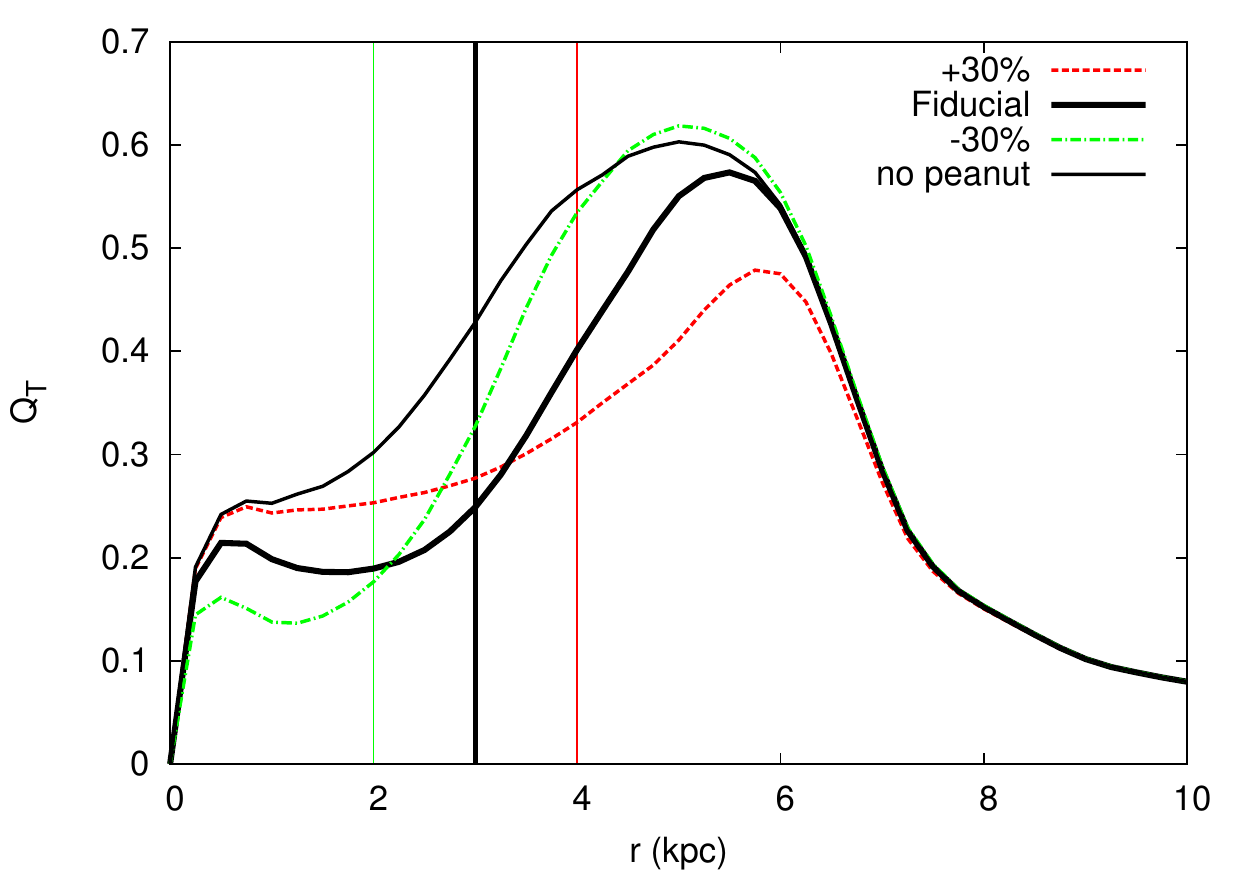}
	\label{fig:Qt_pealength}}
\quad
\caption{\emph{Top Row: Left:} Values of z$_0$ for the fiducial peanut strength (solid thick black line), for 50\% larger (solid thin red line), 25\% larger (dashed green line), 25\% less (dotted magenta line), 50\% less (dot-dashed cyan line) and 0\% peanut strength, i.e. an isothermal sheet (thin solid black line). \emph{Right:} Bar-induced torque $Q_T$ as a function of radius for models with the aforementioned height functions (respective colours).
\emph{Middle Row: Left:} The values of z$_0$ for the fiducial peanut width (solid thick black line), 50\% larger (solid thin red line), 25\% larger (dashed green line), 25\% less (dotted magenta line) and 50\% less peanut width (dot-dashed cyan line). \emph{Right}: Bar-induced torque $Q_T$, for aforementioned models (respective colours). 
\emph{Bottom Row: Left:} The values of z$_0$ for models with the fiducial peanut length (thick solid black line), 30\% longer peanut (dashed red line), 30\% shorter peanut (dashed-dotted green line) and a model without a peanut (i.e. 0\% peanut strength-thin solid black line. \emph{Right:} Bar-induced torque $Q_T$ for the aforementioned models (respective colours).
In all plots, the vertical lines correspond to the positions of the peanut maxima for each respective height function.}
\label{fig:Qt_pean_comps}
\end{figure*}

In this section we examine how both the relative errors of $Q_T$ and those of its maximum value $Q_b$ will be affected by uncertainties in the different parameters of the peanut model. We also introduce a new measure of bar strength, $Q_T^{int}$, which takes into consideration the integrated bar-induced torque, along the entire range of the bar. We do so because even though $Q_b$ remains relatively unchanged when adding a B/P bulge to the model, $Q_T$ over its whole range is significantly affected (see for example Fig.~\ref{fig:Qt_pealength} and Table \ref{table:bp-nobp_strength}), and we wish to have a measure of this change with a singlr number. The bar strength as defined by $Q_T^{int}$ is given by 

\begin{equation}
Q_{T}^{int}=\frac{1}{r_{disc}}\int\limits_0\limits^{\mathrm{r_{bar}}} Q_T \, \mathrm{d}r   ,
\label{eq:Qbint}
\end{equation}

\noindent where $r_{disc}$ is the disc scalelength.

To get a good estimate of the difference of $Q_T$ between two models over the entire radial range, it is best to carry out a point by point comparison, and then consider the radially averaged relative $Q_T$ error.
The relative error of $Q_T^{int}$ is a better proxy for this error than $Q_b$, although there are cases where the relative error of $Q_T^{int}$ is small, while the average relative error of $Q_T$ is much more significant (such as the first row of Table \ref{table:error_qt_length}) or vice-versa (first row of Table \ref{table:error_qt_comb}) . Therefore it is possible to have two cases with identical $Q_T^{int}$, but locally different $Q_T$.


\subsubsection{Peanut strength uncertainties}

We see in Figs.~\ref{fig:z0_peastrength} and \ref{fig:Qt_peastrength} that $Q_T$ increases as we reduce the strength of the peanut, and it reaches its maximum value when the peanut strength is zero, which corresponds to the height function of a flat isothermal sheet. 

The values of the average and maximum relative errors of $Q_T$ ($\langle$Error $Q_T$$\rangle$ and MAX(Error $Q_T$) respectively), as well as the relative error of $Q_b$ and Q$_T^{int}$ can be seen in Table \ref{table:error_qt_strength} (and in all subsequent tables in the following subsections). We see that when we compare an isothermal sheet to the fiducial peanut model the error in $Q_b$ is of the order of 4\%. This is not representative of the large change that the average relative error of $Q_T$ undergoes (27\%). This is due to the fact that the maximum of $Q_T$ does not change much, even though $Q_T$ itself is affected by a significant amount over its entire range (see Fig.~\ref{fig:Qt_peastrength}). On the other hand, the change in Q$_T^{int}$, which takes into account the whole bar region, is more representative of the change in the average relative error of $Q_T$ (20\%). 

In all the cases and for all the measurements of bar strength, the error introduced in the model due to uncertainty in peanut strength is not as large as the error introduced when not including a B/P bulge in the model.

\begin {table}
\caption {Percentage Error of Bar Strength due to Peanut Strength Uncertainty}
\begin{center}
    \begin{tabular}{r | r | r | r | r  }
    \hline
    Peanut & \multirow{2}{*}{$\langle$Error $Q_T$$\rangle$} &  \multirow{2}{*}{MAX(Error $Q_T$)} &  \multirow{2}{*}{$Q_b$} &  \multirow{2}{*}{Q$_T^{int}$} \\
    Strength & & & & \\ \hline
    +50\% & 8\% & 17\% & 1.4\% & 6\% \\ 
    +25\% & 4\% & 9\% & 0.7\% & 3\% \\  
    -25\% & 5\% & 11\% & 0.5\% & 4\% \\ 
    -50\% & 11\% & 26\% & 1.6\% & 7\% \\ 
     no peanut & 27\% & 74\% & 4\% & 20\%  \\ \hline
    \end{tabular}
\end{center}
The error induced in the bar strength due to different amount of error in the peanut strength. We see the effect of these uncertainties on the average and maximum relative error in $Q_T$ ($\langle$Error $Q_T$$\rangle$ and MAX(Error $Q_T$) respectively), as well as on the relative errors of $Q_b$ and $Q_T^{int}$.
\label{table:error_qt_strength}
\end{table}


\subsubsection{Peanut width uncertainties}

We compare setups with varying peanut widths to our fiducial model. Comparisons for $Q_T$ can be seen in Figs.~\ref{fig:z0_peawidth} and \ref{fig:Qt_peawidth} and as previously mentioned, the mismatch between the different models is found when the scaleheights of the models are different. The extent of the region of $Q_T$ which is flattened is reduced when the width of the B/P is reduced, as expected. Conversely, when we increase the width of the B/P bulge, the area of $Q_T$ which is flattened is increased. 

Values for the errors in the bar region are given in Table \ref{table:error_qt_width}. We see that errors in peanut width do not induce very large errors in the average relative error of $Q_T$, compared to the errors induced when not including a B/P bulge. The errors induced in $Q_T^{int}$ are not very large either, although $Q_b$, in the case of +50\% peanut width, has a relative error larger than that of not including a B/P bulge. This again shows the importance of carrying out a point by point comparison, and a comparison of $Q_T^{int}$, in order to determine the errors induced in bar strength due to uncertainties.

\begin {table}
\caption {Percentage Error of Bar Strength due to Peanut Width Uncertainty}
\begin{center}
   \begin{tabular}{r | r | r | r | r }
   \hline
  Peanut & \multirow{2}{*}{$\langle$Error $Q_T$$\rangle$} &  \multirow{2}{*}{MAX(Error $Q_T$)} &  \multirow{2}{*}{$Q_b$} &  \multirow{2}{*}{Q$_T^{int}$} \\
    Width & & & & \\ \hline
    +50\% & 6\% & 12\% & 9\% & 6\% \\ 
    +25\% & 4\% & 11\% & 5\% & 4\% \\  
    -25\% & 7\% & 20\% & 3\% & 6\% \\ 
    -50\% & 16\% & 39\% & 5\% & 13\% \\ 
	no peanut & 27\% & 74\% & 4\% & 20\%  \\ \hline
    \end{tabular}
\end{center}
As in Table \ref{table:error_qt_strength} but for errors in peanut width.
\label{table:error_qt_width}
\end{table}


\subsubsection{Peanut length uncertainties}
\label{sec:qt_comp_pea_length}

The results of this study are shown in Figs.~\ref{fig:z0_pealength} and \ref{fig:Qt_pealength} and in Table~\ref{table:error_qt_length}. Something worth noting in the Fig.~\ref{fig:Qt_pealength} is that the flattening of $Q_T$ occurs at the positions where the maxima of the peanut are found (which are indicated by the corresponding vertical lines).

In Table \ref{table:error_qt_length} we see the errors induced in the different measurements of bar strength due to uncertainties in peanut length. For the case where the peanut length is 30\% larger than the fiducial value, $Q_b$ has a relative error of 17\% compared to the error of 4\% in $Q_b$ when we do not add a peanut. This seems to suggest that it can be counter-productive to include a B/P in the model when there exist uncertainties in peanut length. However, if we examine the average relative error in $Q_T$, we see that the error induced in $Q_T$ is in fact larger when we do not model a B/P than when we miscalculate its length by +30\%. This points once again to the need for examining the average errors of $Q_T$ and not just $Q_b$, as the errors induced in $Q_b$ are not representative of the error induced in $Q_T$.

We also see in Fig.~\ref{fig:Qt_pealength}, that even though $Q_T$ of the two cases (of fiducial peanut and +30\% peanut length) differs significantly point by point, the area under the curve for the two cases is quite similar. This is reflected in the value of the relative error of $Q_T^{int}$,  which only suffers a change of around 5\% while the average error of $Q_T$ suffers a change of 16\%. We see therefore that $Q_T^{int}$ is not always a good approximation for the average relative error of $Q_T$.

The important thing to note is that all the cases considered induce less error in the average error of $Q_T$ than not modelling the B/P at all.

\begin {table}
\caption {Percentage Error of Bar Strength due to Peanut Length Uncertainty}
\begin{center}
    \begin{tabular}{ r | r | r | r | r }
    \hline
    Peanut & \multirow{2}{*}{$\langle$Error $Q_T$$\rangle$} &  \multirow{2}{*}{MAX(Error $Q_T$)} &  \multirow{2}{*}{$Q_b$} &  \multirow{2}{*}{Q$_T^{int}$} \\
    Length & & & & \\ \hline
    +30\% & 16\% & 35\% & 17\% & 5\% \\ 
    +16\% & 9\% & 19\% & 7\% & 3\% \\  
    +8\% & 5\% & 10\% & 3\% & 1\% \\ 
    -8\% & 4\% & 10\% & 2\% & 1\% \\ 
    -16\% & 9\% & 19\% & 4\% & 3\% \\
    -30\% & 16\% & 39\% & 8\% & 8\% \\ 
	no peanut & 27\% & 74\% & 4\% & 20\%  \\ \hline
    \end{tabular}
\end{center}
As in Table \ref{table:error_qt_strength} but for errors in peanut length.
\label{table:error_qt_length}
\end{table}

\subsubsection{Combinations of uncertainties}

\begin{figure*}
\centering
\subfigure{%
	\includegraphics[width=0.42\textwidth]{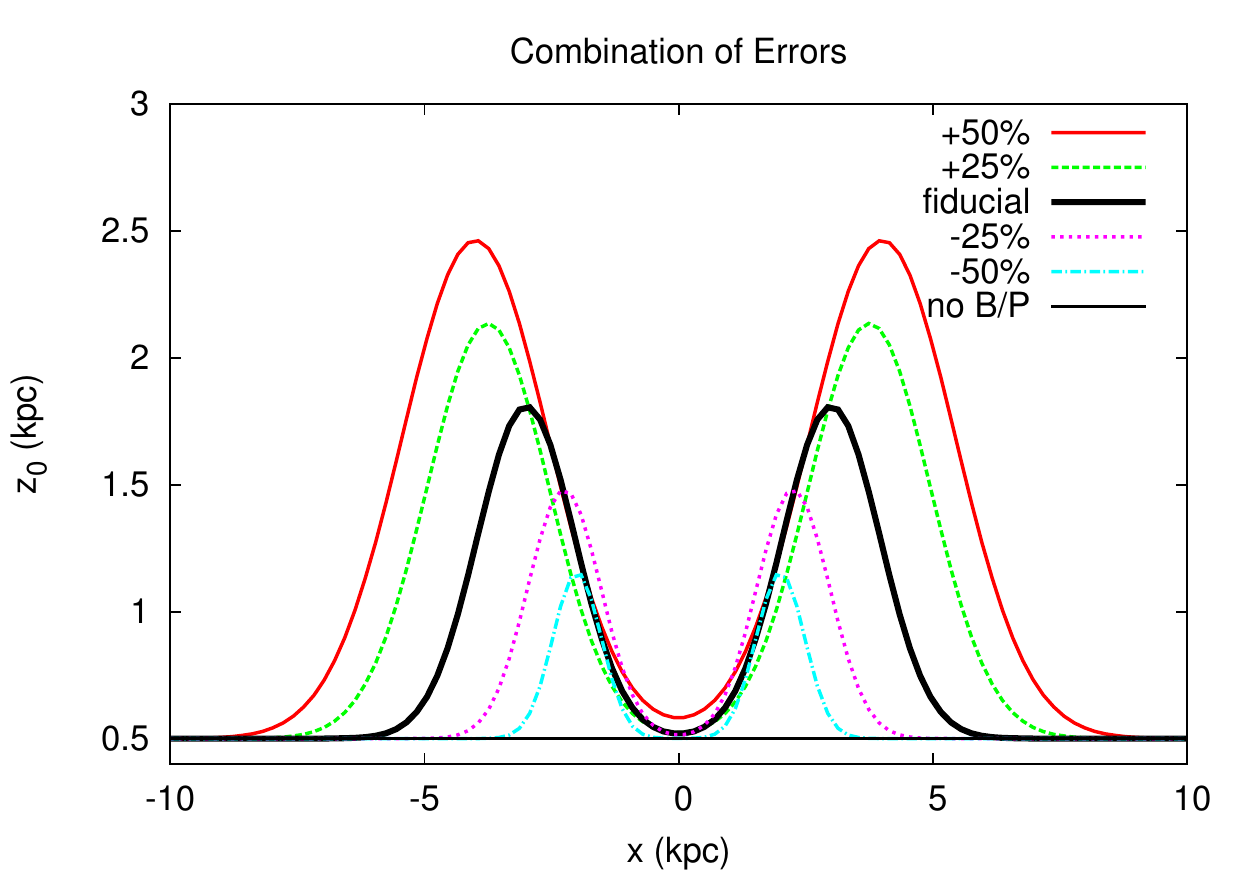}
	\label{fig:z0_peacomb}}
\quad
\subfigure{%
	\includegraphics[width=0.42\textwidth]{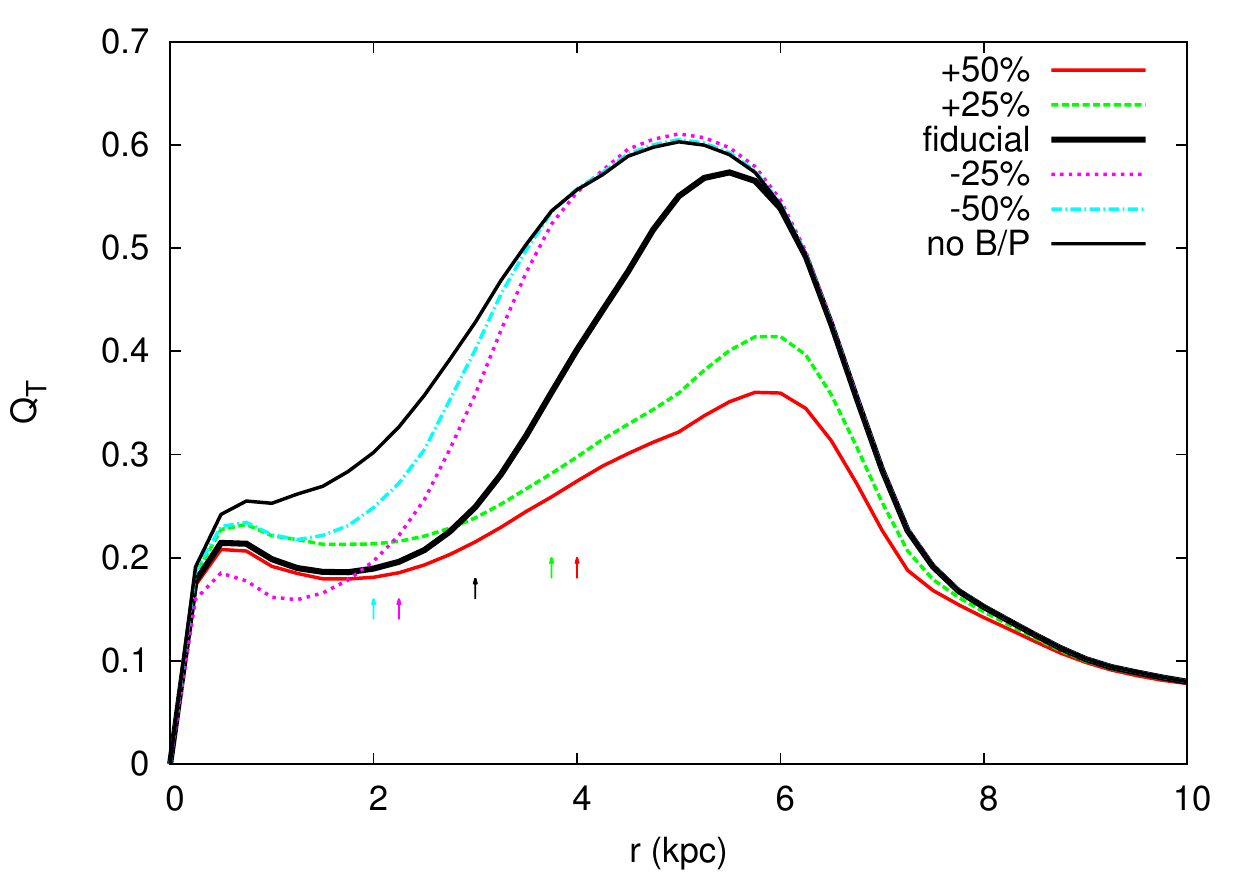}
	\label{fig:Qt_peacomb}}
\quad
\caption{\emph{Left:} Values of z$_0$ for different combinations of uncertainties.
\emph{Right:} Strength of non-axisymmetric forcings $Q_T$ as a function of radius, for the different combinations of uncertainties. In order to not over-clutter the plot, the positions of the peanut maxima are given by the vertical arrows, from left to right, for the -50\%, -25\%, fiducial, +25\% and +50\% cases.}
\label{fig:Qt_pean_combs}
\end{figure*}

As has already been discussed, the most likely scenario is that of a combination of different sources of error affecting our model. The combinations of errors shown in Table \ref{table:error_qt_comb} are as in Section \ref{sec:comb_potfor}. We see that the average and maximum relative error in $Q_T$ for all the combinations is less than that of not modelling the B/P at all. The scaleheights and bar strength for these models can be seen in Figs. \ref{fig:z0_peacomb} and \ref{fig:Qt_peacomb} respectively.

\begin {table}
\caption {Percentage Error of Bar Strength due to a Combination of Uncertainties}
\begin{center}
    \begin{tabular}{ r | r | r | r | r}
    \hline
    All Errors & $\langle$Error $Q_T$$\rangle$ & MAX(Error $Q_T$) & $Q_b$ & Q$_T^{int}$ \\ \hline
    +50\% & 19\% & 42\% & 47\% & 25\% \\ 
    +25\% & 16\% & 35\% & 28\% & 14\% \\  
    -25\% & 16\% & 50\% & 7\% & 11\% \\ 
    -50\% & 20\% & 63\% & 7\% & 17\% \\ 
	no peanut & 27\% &  74\% & 4\% & 20\%  \\ \hline
    \end{tabular}
\end{center}
As in Table \ref{table:error_qt_strength} but for different combinations of uncertainties.
\label{table:error_qt_comb}
\end{table}

\section{\bf{Summary \& Conclusions}}
\label{sec:Summary}

In this paper we present the effects of a Boxy/Peanut height function on the potential, forces, periodic orbits and bar strength of a barred galaxy. We show that such height functions significantly affect the results, which consequently hints to the effects that a Boxy/Peanut bulge will have on its host galaxy. 

We present a method for calculating the potential and forces due to the stellar component of a disc galaxy, based on a three-dimensional integration of the stellar density distribution, which can be obtained from images of not too inclined galaxies combined with a given height function. The method gives robust results for different test cases, as well as allowing for any general height function to be used, thus allowing for complex density distributions to be modelled.  

We used our code on an image extracted from a $N$-body+SPH simulation of an isolated galaxy, together with two flat, position-independent height functions, and two position-dependent height functions. Of the two position-dependent height functions, one models a peanut bulge and one models a boxy bulge. To create an accurate and physically motivated fiducial height function for the peanut, we shaped and fitted our peanut height function to the Boxy/Peanut bulge of the simulated galaxy. 

We found, in accordance with previous results in the literature \citep{HeikiEija2002}, that for the two flat height functions the potential and forces do not vary much, provided the setups have equivalent scaleheights. This also holds true for the bar strength $Q_T$, which does not change much for different flat height functions.

However, we found that for boxy or peanut height functions the potential and forces vary significantly with respect to the case in which a flat height function is used (see Figure \ref{fig:resultsgtr116potforce}). For the potential, the difference can be up to 7\% for an extended region within the bar. For $F_x$ the difference can be as large as 37\%, while for $F_y$ this difference can be as large as 28\%. We therefore concluded that if a Boxy/Peanut bulge is present, one should include it when creating a dynamical model of the galaxy. 

To further confirm this result, we examined the effect of the Boxy/Peanut bulge on the morphology of the most important families of periodic orbits found in barred galaxies. We see that by taking into account the B/P geometry (i.e. by using our fiducial peanut height function) for a given energy, the elongation of the x$_1$ orbits -- the bar-supporting orbits elongated parallel to the bar -- is decreased; this effect is most noticeable for orbits in the region where the Boxy/Peanut is maximum (around $\pm3\,\mathrm{kpc}$, see Figure \ref{fig:difx1orb}) as expected. By adding a Boxy/Peanut to the model the extent of the x$_2$ family -- the family of orbits perpendicular to the bar -- is increased by $\sim$43\% (see Figure \ref{fig:orbgtr116Ej}), as is the distance between the two Inner Lindblad Resonances (ILRs). Additionally, the position of the 3:1 resonance is changed; the 3/1 family -- elongated along the bar and asymmetric with respect to the $y$-axis -- appears at larger energies and is much more extended in the characteristic diagram (see Figure \ref{fig:orbgtr116Ej}). All the aforementioned effects will have an impact on the stellar as well as the gaseous kinematics of the galaxy. The shape and strength of the shocks in the gas will be affected, which in turn affects the amount of gas inflow to the central parts of the galaxy. This could have an impact on the formation of discy bulges and possibly on the fuel reservoir for AGN activity. We plan to investigate in future work the extent of the effects of B/P bulges on gas flows in galaxies.

We also studied the effect of the Boxy/Peanut bulge on the bar strength, as given by the non-axisymmetric forcings due to the bar, $Q_T$. The shape as well as the maximum of $Q_T$ are significantly affected by taking into account the geometry of a Boxy/Peanut bulge. We found it useful to define a new quantity for measuring bar strength, $Q_T^{int}$, which allows us to extract information about the strength of the bar by using its whole extent. The presence of a Boxy/Peanut bulge, especially at the points where its scaleheight is maximum, reduces the bar strength (see Figure \ref{fig:Qt_pea_nopea}) which confirms that the presence of a Boxy/Peanut bulge reduces the bar induced torques.

Even though taking into account the geometry of Boxy/Peanut bulges will affect the model, it is not trivial to obtain their parameters observationally. We therefore examined how much error would be introduced in the results by introducing uncertainties in the Boxy/Peanut parameters.
Each source of error individually (peanut strength, peanut length and peanut width), as well as combinations of the different sources of error, induce errors in the results which in general are considerably less than those induced by not modelling the peanut at all. So, for realistic values of uncertainties in the peanut parameters, the error in including a peanut will be less than the error induced by not including a peanut in the model. 

The simulated galaxy we chose for this study contains a strong bar, corresponding to bar classes 5 and 6 from the \citet{ButaBlock2001} classification. Therefore the results of this study can be straightforwardly and quantitatively applied to real galaxies with similar bar and peanut strength, which account for approximately 20\% of SB galaxies in the local universe. Our results are also qualitatively relevant to all barred galaxies in the secular evolution phase, although for reduced bar and peanut strength the effect of the Boxy/Peanut bulge on the model is also reduced. In this work we have presented an in depth study of the effects of a Boxy/Peanut bulge on its galaxy model, focusing on a particular test case; we plan to present a quantitative statistical study of the effects of these bulges on their host galaxies elsewhere.\newline

\section*{Acknowledgements}
All authors acknowledge financial support to the DAGAL network from the People Programme  (Marie Curie Actions) of the European Union's Seventh Framework Programme FP7/2007-2013/ under REA grant agreement number PITN-GA-2011-289313. EA and AB also acknowledge financial support from the CNES (Centre National d'Etudes Spatiales - France). The simulation analysed here was made using HPC resources from GENCI- TGCC/CINES (Grants 2013 - x2013047098 and 2014 - x2014047098). We would also like to thank Dimitri Gadotti for useful discussions on the properties of barred galaxies and for helpful suggestions on the manuscript and Sergey Rodionov for software assistance.

\bibliographystyle{mn2e}
\bibliography{References}


\appendix

\section{The Analytic Galaxy Model}
\label{sec:appendix}

To model the disc we use a Miyamoto-Nagai density \citep{MiyamotoNagai} which is defined by the potential-density pair,
\begin{equation}
\Phi_{MN} (R,z)= - \frac{GM_D}{\sqrt{R^2+(a+\sqrt{z^2+b^2})^2}},
\label{eq:MNpotential}
\end{equation}
\begin{equation}
\begin{split}
\label{eq:MNdensity}
&\rho_{MN}(R,z) = \\
&\left(\frac{b^2 M_D}{4\pi}\right)\frac{aR^2 +(a+3\sqrt{z^2+b^2})(a+\sqrt{z^2+b^2})^2}{[R^2 +(a+\sqrt{z^2+b^2})^2]^{5/2} (z^2+b^2)^{3/2}} ,
\end{split}
\end{equation}
\noindent where $R$ and $z$ are the cylindrical coordinates, \emph{G} is Newton's gravitational constant, $M$ the total mass of the system and $a$ and $b$ are its characteristic lengths. We set the parameters $a$ and $b$ to 9 and 1.8 respectively, such that we obtain a realistic exponential disc with a scalelength of about $3\,\mathrm{kpc}$ with its mass set to 0.56 times the total mass of the system \citep{Gadotti2011}.\\

The bulge is modelled using a Dehnen sphere \citep{Dehnen1993}, where we set $\gamma$=0.5 in order to obtain a cuspy density distribution. The potential density pair is given by,

\begin{equation}
\rho(r) = \frac{5}{8\pi}\frac{r_B M_B}{\sqrt{r}(r_B+r)^{7/2}},
\label{eq:dehdens}
\end{equation}

\noindent and

\begin{equation}
\Phi(r) = \frac{-2GM_B}{3r_B}(1-(\frac{r}{r+r_B})^{3/2}),
\label{eq:dehpot}
\end{equation}
\noindent where $r_B$ is a characteristic radius of the system.
The mass of the bulge is set to 0.34 the total mass of the model which is a typical value for the bulge mass \citep{Gadotti2011}.

The bar is modelled using a Ferrers ellipsoid (Ferrers 1877), whose density is given by,

\begin{equation}
\rho =
\begin{cases}
  \rho_0(1-m^2)^n & m \leq 1 \\
  0 & m \geq 1
\end{cases}
,
\label{Ferrerdef}
\end{equation}

\noindent where $m^2$ is

\begin{equation}
m^2 = \frac{x^2}{\alpha^2} + \frac{y^2}{\beta^2} + \frac{z^2}{\gamma^2} .
\label{mdef}
\end{equation}

The central density of the bar is given by $\rho_0$, while $n$ sets the decrease in bar density as a function of position and $\alpha$, $\beta$ and $\gamma$ give the sizes of the three semi-principal axes.
The mass of the bar is 0.1 times the total mass, which is again a typical value for real galaxies \citep{Gadotti2011}. The bar's semi-major axis is set to $a$=$5\,\mathrm{kpc}$, with an axial ratio of $a/b$=2.5, and we use the inhomogeneous $n$=1 case. 

When integrating orbits in these potentials, we do so in the rotating frame of reference of the bar, where the bar potential rotates with a pattern speed which is set such that corotation occurs just outside the end of the bar within the range $1.4 \textgreater R_{CR}/R_{bar} \textgreater 1$, where $R_{CR}$ and $R_{bar}$ are the corotation and bar radius respectively (e.g. \citealt{Athanassoula1992b}). The equations of motion in a rotating frame of reference are taken from Chapter 3 of \cite{BT2008}, where the fictitious forces due to rotation are taken into account.

\label{lastpage}

\end{document}